\documentclass[12pt]{article}  
\usepackage{graphicx}

\usepackage{colordvi}

\newcommand{\be}{\begin{equation}}
\newcommand{\ee}{\end{equation}}
\newcommand{\ea}{\end{eqnarray}}
\newcommand{\ba}{\begin{eqnarray}}
\def\ni{\noindent}

\def\[{\left\lbrack}
\def\]{\right\rbrack}

\def\({\left(}
\def\){\right)}

\begin{document}

%\markboth{G. Oliveira-Neto et al.}
%{Ho\v{r}ava-Lifshitz quantum cosmology}

\title{Quantum cosmology of a Ho\v{r}ava-Lifshitz model coupled to radiation}

\author{G. Oliveira-Neto and L. G. Martins\\
Departamento de F\'{\i}sica, \\
Instituto de Ci\^{e}ncias Exatas, \\ 
Universidade Federal de Juiz de Fora,\\
CEP 36036-330 - Juiz de Fora, MG, Brazil.\\
gilneto@fisica.ufjf.br, laysamartinsymail@yahoo.com.br
\and G. A. Monerat\\
Departamento de Modelagem Computacional,\\
Instituto Polit\'{e}cnico,\\
Universidade do Estado do Rio de Janeiro,\\
Rua Bonfim, 25 - Vila Am\'{e}lia - Cep 28.625-570,\\
Nova Friburgo, RJ, Brazil.\\
monerat@uerj.br
\and E. V. Corr\^{e}a Silva\\
Departamento de Matem\'{a}tica, F\'{\i}sica e Computa\c{c}\~{a}o, \\
Faculdade de Tecnologia, \\
Universidade do Estado do Rio de Janeiro,\\
Rodovia Presidente Dutra, Km 298, P\'{o}lo
Industrial,\\
CEP 27537-000, Resende-RJ, Brazil.\\
eduardo.vasquez@pq.cnpq.br}

\maketitle
\clearpage
%\pub{Received (Day Month Year)}{Revised (Day Month Year)}

\begin{abstract}
In the present paper, we canonically quantize an homogeneous and isotropic Ho\v{r}ava-Lifshitz 
cosmological model, with constant positive spatial sections and coupled to radiation. We consider 
the projectable version of that gravitational theory without the detailed balance condition. We use the
ADM formalism to write the gravitational Hamiltonian of the model and the Schutz variational
formalism to write the perfect fluid Hamiltonian. We find the Wheeler-DeWitt equation for the
model, which depends on several parameters. We study the case in which parameter values are chosen so that
the solutions to the Wheeler-DeWitt equation are bounded. 
Initially, we solve it using the {\it Many Worlds} interpretation. Using wavepackets computed with
the solutions to the Wheeler-DeWitt equation, we obtain the scalar factor expected value $\left<a\right>$. 
We show that this quantity oscillates between finite maximum and minimum values and never vanishes. 
Such result indicates that the model is free from singularities, at the quantum level. We reinforce this indication 
by showing that by subtracting one standard deviation unit from the expected value $\left<a\right>$, the latter remains
positive. Then, we use the {\it DeBroglie-Bohm} interpretation. Initially, we compute the Bohm's trajectories for the 
scale factor and show that they never vanish. Then, we show that each trajectory agrees with the corresponding
$\left<a\right>$. Finally, we compute the quantum potential, which helps understanding why the scale factor 
never vanishes.
%\keywords{Ho\v{r}ava-Lifshitz gravity; quantum cosmology; big bang singularity; {\it DeBroglie-Bohm} interpretation}
\end{abstract}

%\ccode{PACS Nos.: 04.40.Nr, 04.50.kd, 04.60.Ds, 04.60.Kz, 98.80.Qc}

\section{Introduction}
\label{introduction}

General relativity is presently the most successful theory of gravitation, because it explains in a precise
way several observational phenomena and also predicts several new ones, that have been confirmed over the years. 
The most recent confirmation was the first detection of gravitational waves \cite{abbott}. The application of general
relativity to cosmology gave rise to a very complete and detailed description of the birth and the evolution
of our Universe. Unfortunately, general relativity is not free of problems. In a series of
theorems it has been shown that, for very general and reasonable conditions, a large class of spacetimes satisfying
the general relativity field equations do develop singularities \cite{hawking}. Those singularities develop
under extreme gravitational conditions and once they appear general relativity loses its predictive power.
One proposal for eliminating those singularities was the quantization of general relativity. Unfortunately,
it was shown that general relativity is not perturbatively renormalizable \cite{weinberg}. After that discovery
many geometrical theories of gravity, distinct from general relativity and perturbatively renormalizable, have 
been introduced. Regrettably, those theories produce massive ghosts in their physical spectrum and they are
not unitary theories\cite{stelle}.

In 2009 Petr Ho\v{r}ava introduced a geometrical theory of gravity with a different property \cite{horava}. In his theory,
nowadays known as Ho\v{r}ava-Lifshitz theory (HL), there is an anisotropic scaling between space and time. His inspiration
came from condensed matter physics where that anisotropy between space and time is common and it is represented by a
dynamical critical exponent $z$ \cite{ma,halperin,sachdev,fradkin}. For physical systems which satisfy Lorentz invariance we have $z=1$.
The main motivation of Ho\v{r}ava for the introduction of that anisotropy is that of improving the short-distance behavior of the theory. It means that 
Lorentz symmetry is broken, at least at high energies, where that 
asymmetry between space and time takes place. At low energies the HL theory tends to GR, thus recovering
Lorentz symmetry. 
As discussed by Ho\v{r}ava \cite{horava}, a theory
of gravity using those ideas is power-counting renormalizable, in 3+1 dimensions, for $z = 3$.  Besides, GR is recovered when $z \to 1$.
The HL theory was formulated, originally, with the aid of the Arnowitt-Deser-Misner (ADM) formalism \cite{misner}.
In the ADM formalism the four  dimensional metric $g_{\mu\nu}$ ($\mu, \nu = 0,1,2,3$) is decomposed in terms
of the three dimensional metric $h_{i j}$ ($i, j = 1,2,3$), of spatial sections, the shift vector $N_i$ and the lapse function ${\cal N}$, which is viewed as a gauge field for time reparametrizations.
In general all those quantities depend both on space and time. In his original work, Ho\v{r}ava considered the simplified assumption that 
${\cal N}$ should depend only on time \cite{horava}. This assumption has became known as the {\it projectable condition}. 
Although many works 
have been written about HL theory using the {\it projectable condition}, some authors have considered the implications of working in the 
{\it non-projectable condition}. In other words, they have considered ${\cal N}$ as a function of space and time \cite{blas,blas1}.
The gravitational action of the HL theory was proposed such that the kinetic component was constructed separately from the potential one.
The kinetic component was motivated by the one coming from GR, written in terms of the extrinsic curvature tensor. It contains
time derivatives of the spatial metric up to the second order and one free parameter ($\lambda$), which is not present in the general relativity 
kinetic component.   At the limit $\lambda \to 1$, one recovers GR kinetic component. The potential component must
depend only on the spatial metric and its spatial derivatives. As a geometrical theory of gravity, the potential component of the HL theory
should be composed of scalar contractions of the Riemann tensor and its spatial derivatives. 

In his original paper \cite{horava}, Ho\v{r}ava considered a simplification in order to reduce the number of possible terms contributing to the potential component of his theory. It is called the {\it detailed balance condition}. Although this condition indeed reduces the number of terms contributing to the potential component,
some authors have shown that, without using this condition, it is possible to construct a well defined and phenomenologically interesting theory, 
without many more extra terms \cite{mattvisser,mattvisser1}. Like other geometrical theories of gravity, it was shown that the projectable 
version of the HL theory, with the detailed balance condition, has massive ghosts and instabilities \cite{mattvisser1,wang}. The HL theory has been applied to cosmology and produced very interesting 
models \cite{bertolami,saa,kord,pedram2,misonoh,ivano,bramberger,gil3}. For a recent review on some aspects of the HL theory, see Ref.\cite{wang1}.

One of the first attempts to quantize the gravitational interaction was the canonical quantization of general relativity (CQGR).
When applied to homogeneous cosmological spacetimes, the CQGR gives rise to quantum cosmology (QC). Although many physicists
believe that QC is not the correct theory to describe the Universe, at the quantum level, an important point has been raised by that
theory. It is related to the interpretation of that quantum theory of the whole Universe. The {\it Copenhagen} interpretation of 
quantum mechanics cannot be applied to that theory because it is not possible to apply a statistical interpretation
to a system composed of the entire Universe. One cannot repeat experiments for that system. Two important interpretations
of quantum mechanics that can be used in QC are those known as the {\it Many Worlds} \cite{everett} and the {\it DeBroglie-Bohm} \cite{bohm,holland} interpretations.
In many aspects they lead to the same results as the {\it Copenhagen} interpretation and can be applied to a system composed of the
entire Universe. The {\it Many Worlds} is the interpretation most commonly used in QC, although the {\it DeBroglie-Bohm} interpretation has been 
applied to several models of quantum cosmology with great success
\cite{gil3,kim,acacio1,pedram1,gil,vakili,das,das1,gil2}. For more references on the use of the {\it DeBroglie-Bohm} interpretation 
in QC see Ref.\cite{julio}. In most of those models, the authors compute the scale factor trajectory
and shows that this quantity never vanishes. That result gives a strong indication that those models are
free from singularities, at the quantum level. Another important quantity introduced by the
{\it DeBroglie-Bohm} interpretation is the quantum potential ($Q$) \cite{bohm,holland}. 
For those quantum cosmological models, the determination of $Q$ helps understanding why the scale factor never vanishes.

Another interesting application of the {\it DeBroglie-Bohm}
interpretation in QC is a recent proof of the idea that the Universe
could be spontaneously created from nothing. In Ref. \cite{he}, the
authors show that a Friedmann-Robertson-Walker (FRW) quantum cosmological	
model, 
without any matter content, produces exponentially growing
Bohm's trajectories for the scale factor, for a particular operator ordering.
The exponential expansion ends when the Universe becomes large
enough such that the early Universe appears. The authors show that such
expansion may be explained by the presence of a specific term in $Q$, for that model, which 
has the same mathematical expression of one that would be produced in the classical potential if a
cosmological constant were present. Finally, in Ref. \cite{he1}, the
authors introduce a new
interpretation for the square modulus of the
wavefunction of the Universe ($\rho$). For a certain FRW quantum
cosmological model, they show that $\rho$, 
which, in that case, is a
function of the scale factor $a$, represents the probability density of
the universe staying in the state where the scale factor assumes the value $a$, during its evolution. 
The authors call it the {\it dynamical interpretation} of the wavefunction of the Universe.
As we shall see, in Section 3, it is not possible to use that interpretation
for the present HL cosmological models.

In the present paper, we canonically quantize a homogeneous and isotropic Ho\v{r}ava-Lifshitz 
cosmological model, with constant positive spatial sections and coupled to radiation. We consider 
the projectable version of that gravitational theory without the detailed balance condition. We use the
ADM formalism to write the gravitational Hamiltonian of the model and the Schutz variational
formalism to write the perfect fluid Hamiltonian. We find the Wheeler-DeWitt equation for the
model. That equation depends on several parameters coming from the HL theory. We study the case in which the values of the 
HL parameters are such that the solutions to the Wheeler-DeWitt equation are bounded. 
Initially, we solve it using the {\it Many Worlds} interpretation. Using wavepackets computed with
the solutions to the Wheeler-DeWitt equation, we obtain the scalar factor expected value $\left<a\right>$. 
We show that this quantity oscillates between maximum and minimum values and never vanishes, indicating that
the model is free from singularities at the quantum level. This indication is further reinforced by 
the observation that if one unit of standard deviation is subtracted from the expected value $\left<a\right>$, what results is still positive.
We also study how the expected value of the scale factor depends on each of the HL parameters. 
Next, now from the standpoint of {\it DeBroglie-Bohm} interpretation, we compute Bohm's trajectories for the 
scale factor, showing that they never vanish. We show that each trajectory agrees with the corresponding
$\left<a\right>$. In addition, we also compute the quantum potential, which helps understanding why the scale factor 
never vanishes. 

It is important to mention that in Refs.\cite{kord,gil3}, the authors studied the QC version of the present 
model with $k \neq 0$, but neglecting the HL parameters $g_C$, $g_\Lambda$ and $g_r$. There,
the {\it Many Worlds} interpretation\cite{kord,gil3} and the {\it DeBroglie-Bohm} interpretation\cite{gil3} were used. In
Ref.\cite{saa}, the authors studied the QC version of the present model with $k=1$, 
using the {\it Many Worlds} interpretation, but neglecting the HL parameter $g_\Lambda$.  
In the present work, we will study the
QC version of the HL model with $k=1$, without neglecting any HL parameter and using both the {\it Many Worlds} and
the {\it DeBroglie-Bohm} interpretations.

Taking into account current cosmological observations, the model introduced here is {\em not}
able to describe the present accelerated expansion of our Universe \cite{expansion}. However, it is not 
our intention to describe the present stage of our Universe with such model; rather, we intend to
describe a `possible' stage of our primordial Universe. Of course, after that initial stage 
the Universe would have to undergo a transition in which the HL parameters should change in order to allow an accelerated expansion.

In Section \ref{classical}, we construct the classical version of the homogeneous and isotropic HL
cosmological model, with constant positive spatial sections and coupled to radiation. In Section \ref{many worlds},
we quantize the classical version of the model and solve the resulting Wheeler-DeWitt equation. Using the solutions,
we construct wavepackets and compute the scale factor expected value, and investigate how the latter
depends on each of the HL parameters. Finally, we evaluate the behavior of the scale factor expected value after 
subtracting from it one unit of standard deviation of $a$. In Section \ref{debroglie-bohm}, we compute Bohm's trajectories 
for the scale factor and the corresponding quantum potentials, also investigating how the Bohm's trajectories depend on the HL parameters. 
We also compare Bohm's trajectories for the scale factor with the corresponding expected values of that quantity.
Section \ref{conclusions} summarizes our main points and results.

\section{Classical Ho\v{r}ava-Lifshitz model coupled to radiation}
\label{classical}

In the present work, we shall consider homogeneous and isotropic spacetimes. They are described by the FRW line element, given by
\begin{equation}
\label{1}
ds^2 = - {\cal N}(t)^2 dt^2 + a(t)^2\left( \frac{dr^2}{1 - kr^2} + r^2 d\Omega^2
\right)\, ,
\end{equation}
\ni in which $d\Omega^2$ is the line element of the two-dimensional sphere with
unitary radius, $a(t)$ is the scale factor, ${\cal N}(t)$ is the lapse function \cite{wheeler} and $k$ represents the
constant curvature of the spatial sections. The curvature is positive for 
$k=1$, negative for $k=-1$ and zero for $k=0$. Here, we are using the natural
unit system, in which $c = 8\pi G = \hbar = 1$.  We assume that the matter content of the model is
represented by a perfect fluid with four-velocity $U^\mu = \delta^{\mu}_0$
in the co-moving coordinate system used. The energy-momentum tensor is given by
\begin{equation}
T_{\mu\nu} = (\rho+p)U_{\mu}U_{\nu} + p g_{\mu\nu} ,
\label{2}
\end{equation}
in which $\rho$ and $p$ are the energy density and pressure of the fluid,
respectively. The Greek indices $\mu$ and $\nu$ run from zero to three. The equation of state for a perfect fluid is $p = \omega\rho$, 
in which $\omega$ is a constant the value of which specifies the type of fluid.

The action for the projectable HL gravity, without the detailed balance condition, for $z=3$ and in $3+1$-dimensions is given by \cite{bertolami},
\begin{eqnarray} \label{3}
\mathcal{S}_{HL} & = & \frac{M_{p}^{2}}{2} \int d^{3}x dt {\cal N} \sqrt{h}  \left[ K_{ij}K^{ij} - \lambda K^{2} - g_{0}{M_{p}}^{2} - g_{1} R - {M_{p}}^{-2}\Big( g_{2}R^{2} + g_{3}R_{ij}R^{ij} \Big) \right. \nonumber \\
& - &  \left. {M_{p}}^{-4} \left( g_{4}R^{3} + g_{5}R R^{\,i}_{\phantom{i}j} R^{\,j}_{\phantom{j}i} + g_{6} R^{\,i}_{\phantom{i}j} R^{\,j}_{\phantom{j}k} R^{\,k}_{\phantom{k}i} + g_{7} R \nabla^{2} R + g_{8} \nabla_{i}R_{jk} \nabla^{i}R^{jk} \right) \right], \nonumber \\
%& + & M_{p}^{2} \int_{\partial \mathcal{M}}^{} d^{3}x \sqrt{h} K.
\end{eqnarray}
\ni in which $g_i$ and $\lambda$ are parameters associated with HL gravity, $M_p$ is the Planck mass,
$K_{ij}$ are the components of the extrinsic curvature tensor and $K$ represents its trace,
$R_{ij}$ are the components of the Ricci tensor and $R$ is the Ricci scalar, both should be computed with the metric of the
spatial sections $h_{ij}$, $h$ is the determinant of $h_{ij}$ and $\nabla_i$ represents covariant derivatives. The Latin
indices $i$ and $j$ run from one to three. As we have mentioned above, the GR kinetic component is recovered in the limit $\lambda \to 1$.

Introducing the metric of the spatial sections that comes from the FRW space-time (\ref{1}), in the action (\ref{3}) and choosing $g_{0} M_{p}^{2} = 2 \Lambda$ and 
$g_{1} = -1$, we can write the action as
\begin{eqnarray}
\label{4}
\mathcal{S}_{HL} & = & \kappa \int_{}^{} dt {\cal N} \left[ - \frac{\dot{a}^{2} a}{{\cal N}^{2}} + \frac{1}{3 \left(3 \lambda - 1 \right)} \left( 6k a - 2\Lambda a^{3} - \frac{12k^{2}}{a M_{p}^{2}} \left( 3g_{2} + g_{3} \right) \right. \right. \nonumber \\
& - & \left. \left. \frac{24k^{3}}{a^{3} {M_{p}^{4}}} \left(9g_{4} + 3g_{5} + g_{6} \right) \right) \right],
\end{eqnarray}
in which $$\kappa = \frac{3(3\lambda-1) M_{p}^{2} }{2} \int_{}^{} d^{3}x \frac{r^{2} \sin \theta}{\sqrt{1 - kr^{2}}}\,\,.$$ If we choose, for simplicity, $\kappa=1$, we will write the HL Lagrangian density ($\mathcal{L}_{HL}$), from $\mathcal{S}_{HL}$ in Eq. (\ref{4}) as,
\begin{equation}
\label{5}
\mathcal{L}_{HL} = {\cal N} \left[ -  \frac{\dot{a}^{2}a}{{\cal N}^{2}} + g_{c}ka - g_{\Lambda} a^{3} - g_{r}\frac{k^{2}}{a} - g_{s} \frac{k^{3}}{a^{3}} \right],
\end{equation}
in which the new parameters are defined by
\begin{eqnarray}
\label{6}
g_{c} & = & \frac{2}{3 \lambda - 1}, \quad
g_{\Lambda} = \frac{2 \Lambda}{3 \left( 3 \lambda - 1 \right)}, \quad
g_{r} = \frac{4}{(3\lambda-1)M_p^2} \left( 3g_{2} + g_{3} \right), \\\nonumber
g_{s} & = & \frac{8}{(3\lambda-1)M_p^4} \left(9 g_{4} + 3g_{5} + g_{6} \right).
\end{eqnarray}
The parameter $g_c$ is positive, by definition, and the others may be either positive or negative.

Now, we want to write the HL Hamiltonian density. To accomplish the task, we must
compute the momentum canonically conjugated to the single dynamical variable present in the geometry sector, i.e., the scale factor. Using the definition, that momentum ($P_a$) is
given by,
\begin{equation}
\label{7}
P_{a} = \frac{\partial \mathcal{L}}{\partial \dot{a}} = \frac{\partial}{\partial \dot{a}} \left[ - \frac{\dot{a}^{2}a}{{\cal N}} \right] = - \frac{2 \dot{a} a}{{\cal N}}\,\,.
\end{equation}
Introducing $P_a$ (\ref{7}) into the definition of the Hamiltonian density, with the aid of $\mathcal{L}_{HL}$ (\ref{5}), we obtain the following HL Hamiltonian ($H_{HL}$):
\begin{equation}
H_{HL}={\cal N}\mathcal{H}_{HL} = {\cal N} \left[ - \frac{P_{a}^{2}}{4 a} - g_{c}ka + g_{\Lambda} a^{3} + g_{r}\frac{k^{2}}{a} + g_{s} \frac{k^{3}}{a^{3}} \right]. 
\label{8}
\end{equation}

In this work, we will obtain the perfect fluid Hamiltonian ($H_{pf}$) using
Schutz's variational formalism \cite{schutz,schutz1}, in which the four-velocity ($U_\nu$) of the 
fluid is expressed in terms of six thermodynamical potentials ($\mu$, $\epsilon$, $\zeta$, $\beta$, $\theta$, $S$), in the following way,
\be
\label{9}
U_\nu = \frac{1}{\mu}\left(\epsilon_{,\nu}+\zeta\beta_{,\nu}+\theta S_{,\nu}\right)\,\,,
\ee
in which $\mu$ is the specific enthalpy, $S$ is the specific entropy. 
The parameters $\zeta$ and $\beta$, absent from the FRW models, are connected to rotation. 
The remaining parameters $\epsilon$ and $\theta$ have no clear physical meaning. The four-velocity obeys the normalization condition,
\be
\label{10}
U^\nu U_\nu = -1.
\ee
The starting point for writing the $H_{pf}$ for the perfect fluid is the action ($\mathcal{S}_{pf}$), which in this formalism is written as
\be
\label{11}
\mathcal{S}_{pf} = \int d^4x\sqrt{-g}(16\pi p)\,\,,
\ee
in which $g$ is the determinant of the four-dimensional metric ($g_{\alpha \beta}$) and $p$ is the fluid pressure. 
Inserting the metric (\ref{1}), Eqs. (\ref{9}) and (\ref{10}), the fluid equation of state and the first 
law of thermodynamics into the action (\ref{11}), and after some thermodynamical considerations, that action takes 
the form \cite{nivaldo},
\be
\label{12}
\mathcal{S}_{pf}=\int dt\left[ 
{\cal N}^{-\frac{1}{\omega}}a^3\frac{\omega(\dot{\epsilon}+\theta\dot{S})^{1+\frac{1}{\omega}}}{(\omega+1)^{1+
\frac{1}{\omega}}}e^{-\frac{S}{\omega}}\right].
\ee
\ni From this action, we can obtain the perfect fluid Lagrangian density and write the Hamiltonian ($H_{pf}$),
\be
\label{13}
H_{pf}={\cal N}{\mathcal{H}_{pf}}={\cal N}\left(P_{\epsilon}^{\omega+1}a^{-3\omega}e^S\right),
\ee
in which $$P_\epsilon = {\cal N}^{-\frac{1}{\omega}}a^3(\dot{\epsilon}+\theta\dot{S})^{\frac{(\omega+1)^{-\frac{1}{\omega}}}{\omega}}e^{-\frac{S}{\omega}}\,\,.$$
We can further simplify the Hamiltonian (\ref{13}), by performing the following canonical transformations \cite{rubakov},
\be
\label{14}
T = -P_S e^{-S}P_\epsilon^{-(\omega+1)},\quad P_T = P_\epsilon^{\omega+1}e^S,\quad \bar{\epsilon} = \epsilon-(\omega+1)\frac{P_S}{P_\epsilon},\quad \bar{P_\epsilon} = P_\epsilon,
\ee
in which $P_S = \theta P_\epsilon$. 
Under these transformations the Hamiltonian (\ref{13}) takes the form
\begin{equation}
H_{pf}={\cal N} {\mathcal{H}_{pf}}= {\cal N}\frac{P_T}{a^{3\omega}},  
\label{15}
\end{equation}
in which $P_{T}$ is the momentum canonically conjugated to $T$. We can write now
the total Hamiltonian of the model ($H$), which is written as the sum of $H_{HL}$ (\ref{8}) with $H_{pf}$ (\ref{15}),
\begin{equation}
H = {\cal N}{\mathcal{H}} = {\cal N} \left[ - \frac{P_{a}^{2}}{4 a} - g_{c}a + g_{\Lambda} a^{3} + \frac{g_r}{a} + \frac{g_s}{a^{3}} + \frac{P_{T}}{a} \right]. 
\label{16}
\end{equation}
Here, we have set $k=1$ (in order to consider only spacelike hypersurfaces with positive constant curvatures)
and $\omega=1/3$ (to restrict the matter content of the Universe to radiation).
The classical dynamics is governed by Hamilton's equations, derived from eq. (\ref{16}).

In order to have an idea of the scale factor classical behavior, we derive Friedmann equation 
by varying $H$ (\ref{16}) with respect to ${\cal N}$ and equating it to zero. In the
ADM formalism,  such equation is also known as the superHamiltonian constraint \cite{wheeler}. Now, in the conformal
gauge ${\cal N}=a$, we have $P_a=-2\dot{a}$, in which the dot means derivative with respect to the conformal time. Therefore, we
may write Friedmann equation in terms of $\dot{a}$ as
\be
\label{17}
\dot{a}^{2}+V_{c}(a)=0,
\ee
in which,
\be
\label{18}
V_{c}(a)=g_{c}a^2-g_{\Lambda}a^{4}-g_{r}-\frac{g_{s}}{a^2}-P_{T}
\ee
is the classical potential. 
The scale factor behavior depends on the particular shape of the classical potential which, in its turn,
depends on the values of its parameters. The parameters $g_c$ and $P_T$ are both positives because $g_c$
is associated to the curvature coupling constant and $P_T$ to the fluid energy density. 
The other parameters may be either positive or negative. In the present work, we shall study the models in which $g_r$ is
positive and $g_s$ and $g_\Lambda$ are negative. For those choices, the scale factor turns out be bounded 
(in other words, it oscillates between maximum and minimum finite values). Those models are thus free from the big bang singularity.

\section{Many Worlds Interpretation}
\label{many worlds}

\subsection{Eigenvalue equation and the spectral method}

We wish to quantize the model following the Dirac formalism for
constrained systems \cite{dirac}. First we introduce a wave-function $\Psi$  which
is a complex function of the canonical variables $a$ and $T$,
\begin{equation}  
\label{19}
\Psi\, =\, \Psi(a ,T)\, .
\end{equation}
By setting up the correspondence between the real variables 
$a$ and $T$ and operators $\hat{a}$ and $\hat{T}$, respectively,        
we then impose appropriate commutators between those operators 
and their corresponding conjugate momenta $\hat{P}_a$ and $\hat{P}_T$.
In the Schr\"{o}dinger picture, the effect of applying $\hat{a}$ and $\hat{T}$ on $\Psi$ amounts
to multiplying $\Psi$ by $a$ and $T$, respectively, whereas operating the conjugate momenta 
on $\Psi$ amounts to applying the differential operators 
\begin{equation}
\hat{P}_{a}\rightarrow -i\frac{\partial}{\partial a}\hspace{0.2cm},\hspace{0.2cm} 
\hspace{0.2cm}\hat{P}_{T}\rightarrow -i\frac{\partial}{\partial T}\hspace{0.2cm}
\label{20}
\end{equation}
on $\Psi$. Finally, we demand that the operator corresponding to $H$ (\ref{16}) ($\hat{H}$) 
annihilate the wave-function $\Psi$. That leads to Wheeler-DeWitt
equation,
\begin{eqnarray}
\bigg(-\frac{1}{4}\frac{\partial^{2}}{\partial a^{2}}+g_{c}a^2-g_{\Lambda}a^{4}-g_{r}-
\frac{g_{s}}{a^{2}}\bigg)\psi(a,\tau) = i\frac{\partial}{\partial \tau}\psi(a,\tau). 
\label{21} 
\end{eqnarray}
in which the new variable $\tau= -T$ has been introduced. 
The operator $\hat{H}$ is self-adjoint \cite{lemos} with respect
to the internal product of two functions $\phi_i$ and $\phi_j$,
\begin{equation}
(\phi_i ,\phi_j ) = \int_0^{\infty} da\, \,\phi_i(a,\tau)^*\, \phi_j (a,\tau)\, ,
\label{22}
\end{equation}
if the wave functions $\phi$ are restricted to those satisfying either
$\phi(0,\tau )=0$ or $\phi^{\prime}(0, \tau)=0$. Here, the prime $\prime$
means the partial derivative with respect to $a$. We consider wave 
functions satisfying the first type of boundary condition and we also 
demand that they vanish when $a \rightarrow \infty$.

The Wheeler-DeWitt equation (\ref{21}) may be solved by writing the wave function $\Psi(a, \tau)$ as
\begin{equation}
\Psi (a,\tau) = e^{-iE\tau}\,\eta(a) , 
\label{23}
\end{equation}
in which $\eta(a)$ depends solely on $a$ and satisfies the eigenvalue equation
\begin{equation}
-\frac{d^2{\eta(a)}}{da^2} + V (a)\,\eta(a)= 4\,E\,\eta(a)\, ,
\label{24}
\end{equation}
so that
\begin{equation}
V(a)=4g_{c}a^{2}-4g_{\Lambda}a^{4}-4g_{r}-\frac{4g_{s}}{a^{2}}.
\label{25}
\end{equation}
In the same way as in the classical regime, the potential $V(a)$
gives rise to bound states. Therefore, the possible values of the energy $E$ in Eq.(\ref{24})
of those states belong to a discrete set of eigenvalues $E_n$, in which $n\in\{1,2,3,...\}$. 
For each eigenvalue $E_n$, there is a corresponding eigenvector $\eta_n(a)$. The general
solution to the Wheeler-DeWitt equation (\ref{21}) is a linear combination of all
those eigenvectors,
\begin{equation}
\label{26}
\Psi(a,\tau) = \sum_{n=1}^{\infty} C_n\eta_n(a)e^{-iE_n\tau},
\end{equation}
in which $C_n$ are constant coefficients to be specified.
We will use Galerkin spectral method (SM) \cite{boyd} in order to solve the 
eigenvalue equation (\ref{24}). This method has already been used in quantum 
cosmology \cite{pedram,gil1,eduardo} and also in several areas of classical general 
relativity \cite{ivano2,henrique,henrique1,henrique2}. 
One important condition for the SM is that the solutions of the 
equation in question must fall sufficiently fast for large values of the independent variable. In the present
situation that variable is the scale factor $a$. Taking into account such restrictions, we impose that 
 $0 < a < L$, in which $L$ is a real number to be suitably chosen.
As we have mentioned above, we shall consider, here, wavefunctions satisfying 
the condition $\Psi (0,\tau)=0$. It is convenient, then, to choose our basis functions to be sine
functions. Therefore, we may write $\eta_n(a)$ in Eq. (\ref{24}) as,
\begin{equation}
\label{27}
\eta(a) \approx \sum_{n=1}^{N} A_n \sqrt{\frac{2}{L}}\sin 
\left({\frac{n\pi a}{L}}\right),
\end{equation}
in which the coefficients $A_n$ are yet to be determined, and a finite number $N$ of base functions has been chosen. 
For the same domain of $a$, we also use the same basis to expand the terms of Eq. (\ref{24}),
\begin{equation}
\label{28}
V(a)\eta(a) \approx \sum_{n=1}^{N} \sum_{m=1}^{N} C_{m,n} A_m \sqrt{\frac{2}{L}}\sin \left({\frac{n\pi a}{L}}\right),
\end{equation}
\begin{equation}
\label{29}
4\eta(a) \approx \sum_{n=1}^{N} \sum_{m=1}^{N} C_{m,n}' A_m \sqrt{\frac{2}{L}}\sin \left({\frac{n\pi a}{L}}\right),
\end{equation}
in which $V(a)$ is given by Eq. (\ref{25}) and the coefficients $C_{m,n}$ and $C_{m,n}'$ can be easily determined to be,
\begin{equation}
\label{32}
C_{m,n} = \frac{2}{L} \int_0^L \sin \left(\frac{m\pi a}{L}\right) V(a)
\sin \left(\frac{n\pi a}{L}\right) da,
\end{equation}
\begin{equation}
\label{33}
C_{m,n}' = \frac{2}{L} \int_0^L \sin \left(\frac{m\pi a}{L}\right) 
4\sin \left(\frac{n\pi a}{L}\right) da.
\end{equation}
Introducing the results in 	Eqs. (\ref{27})-(\ref{33}) into the eigenvalue equation 
(\ref{24}) and due to orthonormality of the basis functions, we obtain
\begin{equation}
\label{34}
\left(\frac{n\pi}{L}\right)^2 A_n + \sum_{m=1}^{N} C_{m,n} A_m =
E \sum_{m=1}^{N} C_{m,n}' A_m,
\end{equation} 
which may be written in compact notation as
\begin{equation}
\label{35}
C'^{-1}\, D\, A\, = E\, A\, ,
\end{equation}
in which $C'$ is the $N \times N$ square matrix the elements of which are given by (\ref{33})
and $D$ is a $N \times N$ square matrix with elements of the form,
\begin{equation}
\label{36}
D_{m,n} = \left(\frac{n\pi}{L}\right)^2 \delta_{m,n} + C_{m,n}.
\end{equation}
The solution to Eq. (\ref{35}) gives the eigenvalues and 
corresponding eigenfunctions to the bound states of our quantum cosmological model.

For the sake of completeness, it is interesting to notice that
the dynamical interpretation of the wavefunction of the Universe,
introduced in Ref. [38], cannot be used in our paper, because Eq. (14),
page 362, of Ref. [38] is {\it not} satisfied here. Although we consider
HL quantum cosmological models and Schutz variational formalism, in
order to describe the matter content of those models, it is
possible to introduce a conserved current $j^a$, from Eq. (24), page 11,
similar to the one given in Eq. (7), page 362, of Ref. [38] and
we can, also, write an expression similar to Eq. (13), page 362, of Ref.
[38]. The problem is that the operator ordering parameter $p$,
introduced in Eq. (5), page 362, of Ref. [38], is $p=0$ in our paper,
from the Wheeler-DeWitt Eq. (21) or Eq. (24), page 11. Therefore, Eq.
(14) of Ref. [38] cannot be satisfied, and the relation
between $\rho(a)$ and the Hubble parameter is not well-behaved, for the
entire domain of $a$.

\subsection{Scale factor expected values and standard deviation}

In the present subsection we solve the eigenvalue equation (\ref{24}) using the SM. In order to understand
how the expected value of the scale factor $\left<a\right>$ depends on each HL parameter, we compute $\left<a\right>$ by
fixing all parameters but one, thus observing how $\left<a\right>$ depends on that varying parameter. 
The procedure is repeated, having all HL parameters vary in the same manner.
We also study how $\left<a\right>$ depends
on the number of base functions $N$ present in $\eta_n(a)$ (\ref{27}). We consider the cases where $2 \leq N \leq 10$.
An important ingredient in the SM is the variable $L$. In order to improve our results, we compute the best value of $L$ for each
value of a given HL parameter.

We compute the scale factor expected value as
\begin{equation}
\label{37}
\left<a\right> = \frac{\int_{0}^{\infty}a\,|\Psi (a,\tau)|^2 da} {
\int_{0}^{\infty}|\Psi (a,\tau)|^2 da},
\end{equation}
in which $\Psi (a,\tau)$ is given by Eq. (\ref{26}). In order to compute $\Psi (a,\tau)$, we use the eigenvectors $\eta_n(a)$ and eigenvalues
$E_n$, derived from Eq. (\ref{24}) with the SM. We, also, set all coefficients $C_n$ to unit in Eq. (\ref{26}).

As we shall see, for all HL parameters and $N$ values considered, $\left<a\right>$ oscillates between maximum and minimum values and never vanishes, hence giving an initial
indication that those models are free from singularities, at the quantum level.
The domain where $\left<a\right>$ oscillates depends on the mean energy ($\bar{E}$) of the wavepacket considered. That quantity is specified by the number $N$, of base functions contributing to the wavepacket, and the energy eigenvalues of those $N$ base functions.
We may refine the evidence that $\left<a\right>$ never vanishes by computing $\underline{a} \equiv \left<a\right>-\sigma_a$,
in which $\sigma_a$ is the standard deviation of $a$. If $\underline a$ is always positive like $\left<a\right>$, it will be a
stronger indication that the model is free from singularities, at the quantum level. We compute $\underline a$, for the present model.
The standard deviation of $a$ is defined as
\begin{equation}
\label{38}
\sigma_a = \sqrt{\left<a^2\right> - \left<a\right>^2},
\end{equation}
in which,
\begin{equation}
\label{39}
\left<a^2\right> = \frac{\int_{0}^{\infty}a^2\,|\Psi (a,\tau)|^2 da} {
\int_{0}^{\infty}|\Psi (a,\tau)|^2 da},
\end{equation}
and $\left<a\right>^2$ is obtained  by squaring Eq. (\ref{37}). By using the wavefunction (\ref{26}) and repeating the same procedure  for computing
$\left<a\right>$, we obtained $\underline a = \left<a\right>-\sigma_a$ for all HL parameters and $N$ values considered. In what follows we present our results
on how $\left<a\right>$ depends on the HL parameters ($g_c$, $g_r$, $g_s$, $g_\Lambda$) and $N$. The values of the HL parameters and $N$, in all figures
below, were chosen for the sake of better visualization of results.

\subsubsection{Behavior of $\left< a \right>$ as $g_c$ varies}

If we fix $N$ and the HL parameters but $g_c$, and let $g_c$ increase, we observe that:
($i$) the maximum value of $\left<a\right>$ decreases;
($ii$) the amplitude of oscillation of $\left<a\right>$ decreases;
($iii$) the number of oscillations of $\left<a\right>$, for a fixed interval of $\tau$, increases.

That behavior may be understood by the observation of the potential that confines the scale factor. As $g_c$ increases, 	
$\left<a\right>$ is forced to oscillate within an ever smaller region. Under those conditions, for fixed $N$ and the other HL parameters, the maximum 
value and the amplitude of $\left<a\right>$ both decrease. Moreover, since the domain where $\left<a\right>$ oscillates is smaller,
the number of oscillations of $\left<a\right>$  for a fixed $\tau$ interval also increases. 
Figs. 1 and 2 illustrate the behavior of the potential $V(a)$ and of the expected value $\left<a\right>$, for two different values of $g_c$
whereas the $\tau$ interval, $N$ and the other HL parameters remain fixed.

\begin{figure}[h!]
\centering
%\subfloat[$V(a)\times a$ para $g_{c}=10$.]{
\includegraphics[scale=0.3]{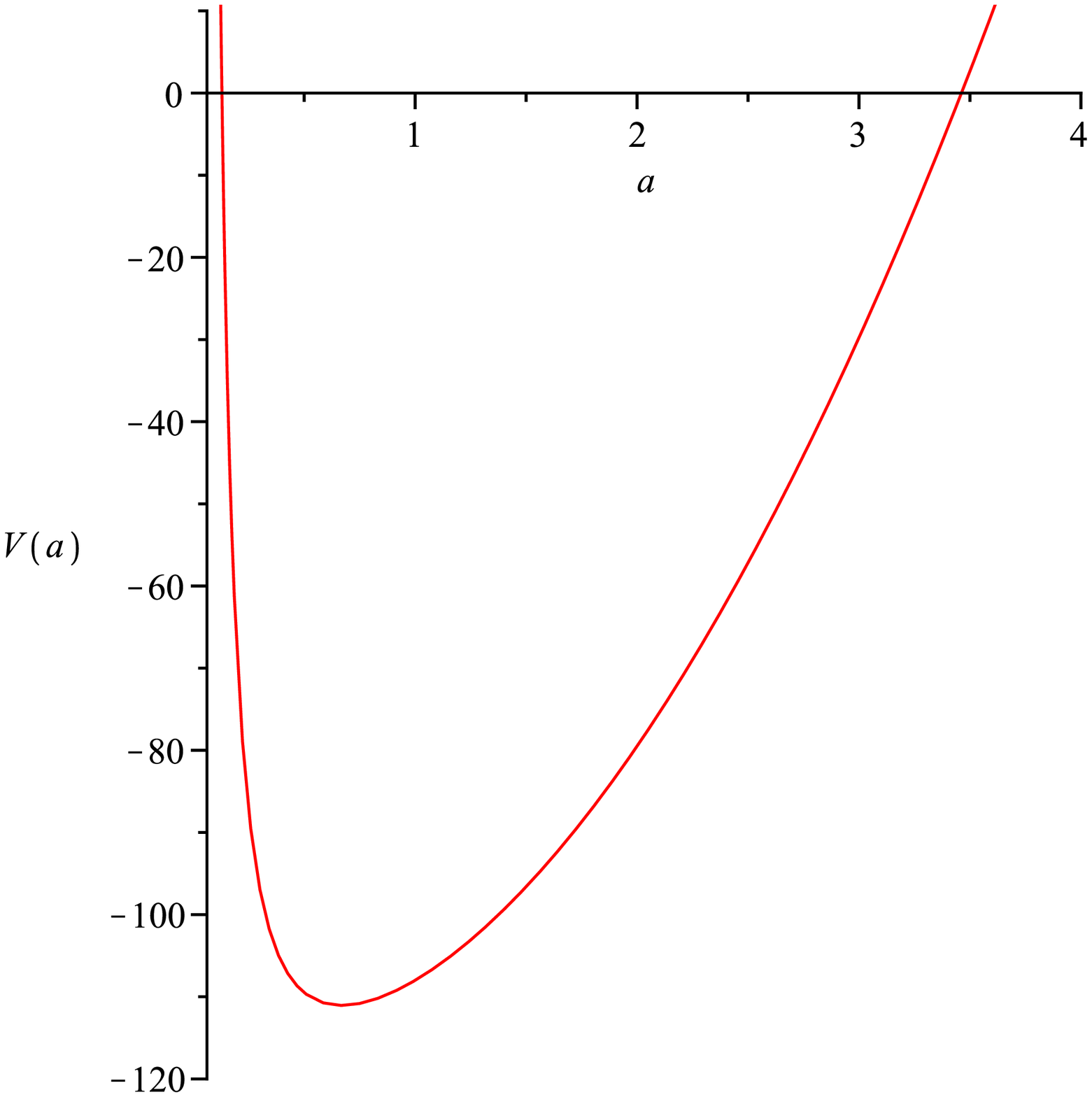}%Gc=10V(a)
\label{fig1a}
%}
\quad
%\subfloat[$V(a)\times a$ para $g_{c}=90$.]{
\includegraphics[scale=0.3]{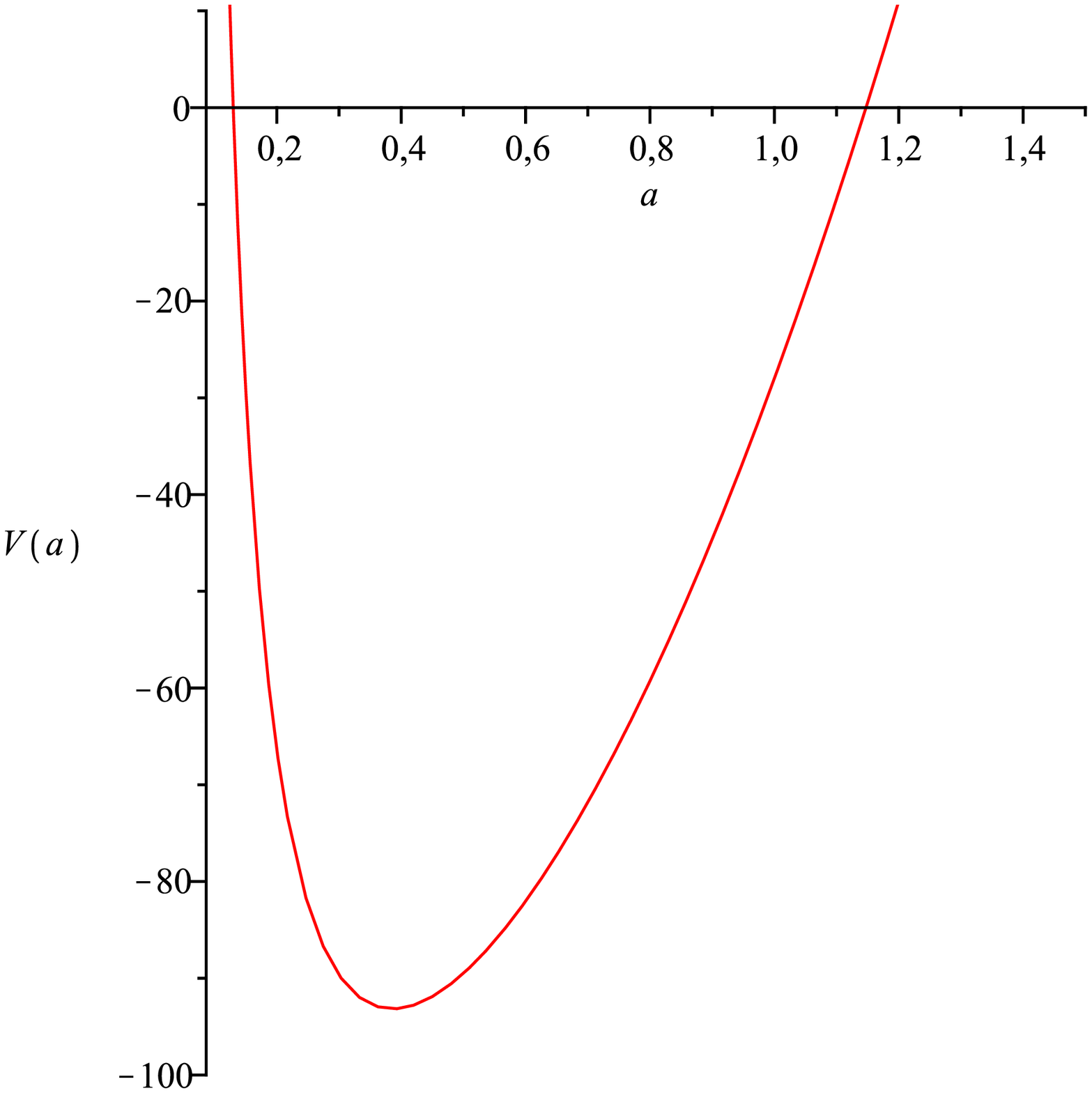}%Gc=90V(a)
\label{fig1b}
%}
\caption{The potential $V(a)$ for $g_s=-2$, $g_r=120$, $g_\Lambda=0$, for 
$g_{c}=10$ (left) and $g_{c}=90$ (right).}
\label{fig1}
\end{figure}

For $g_c = 10$, we have $\bar{E}=-104.592479205836$. Therefore, from the potential $V(a)$ (Fig. 1), the interval where $\left<a\right>$ oscillates is\\
$\[0.3782813789,1.182224715\]$, which has an amplitude of 0.8039433361. On the other hand, for $g_c = 90$, we have $\bar{E}=-73.7774376335110$. 
The interval where $\left<a\right>$ oscillates is $\[0.2184008560,0.6825577576\]$, with smaller amplitude, equal to 0.4641569016.
\begin{figure}[h!]
\centering
%\subfloat[$\langle a \rangle\times t$ para $g_{c}=10$.]{
\includegraphics[scale=0.3]{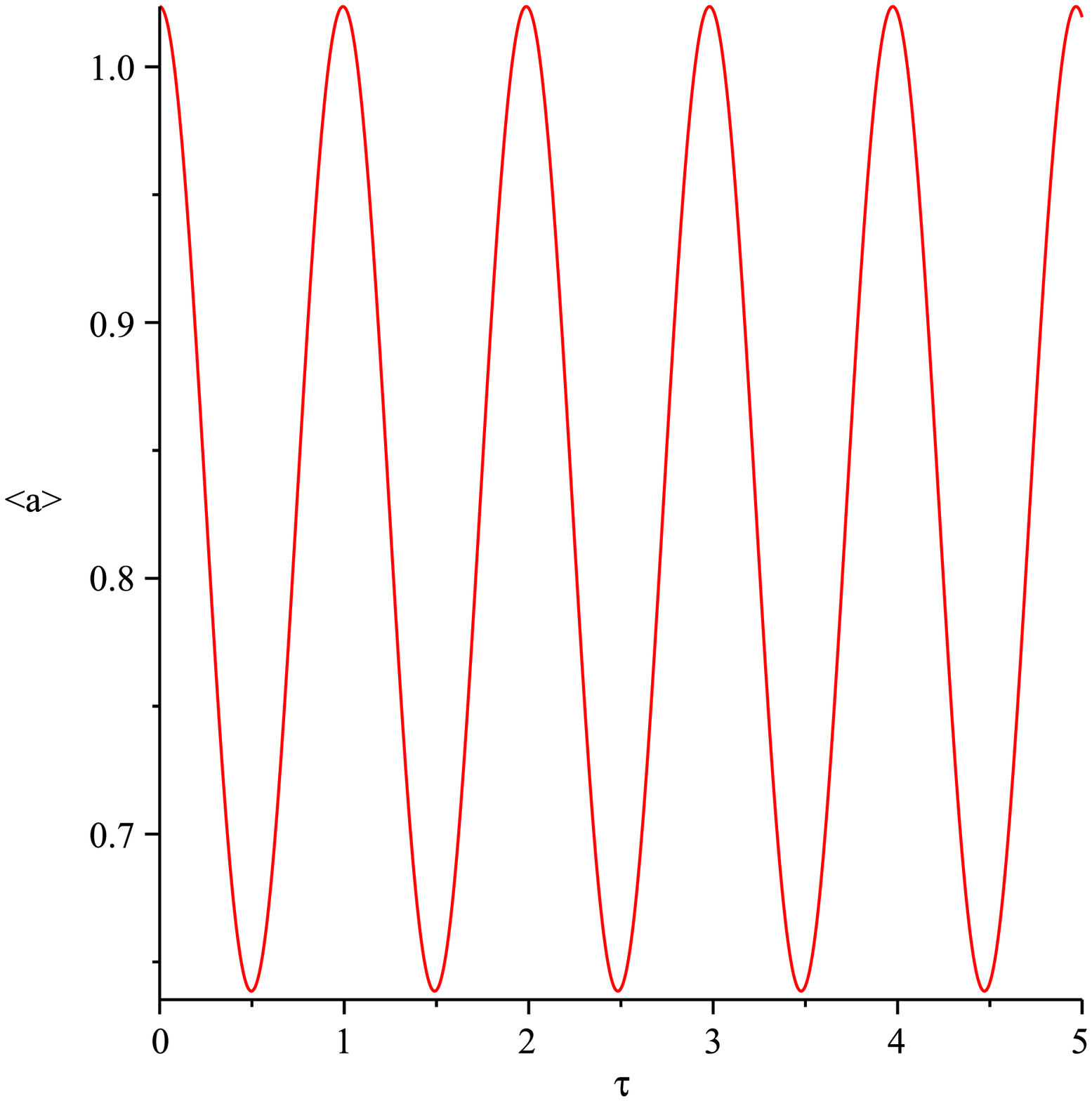}%Gc=10-amedio-N2-L324
\label{fig2a}
%}
\quad
%\subfloat[$\langle a \rangle\times t$ para $g_{c}=20$.]{
\includegraphics[scale=0.3]{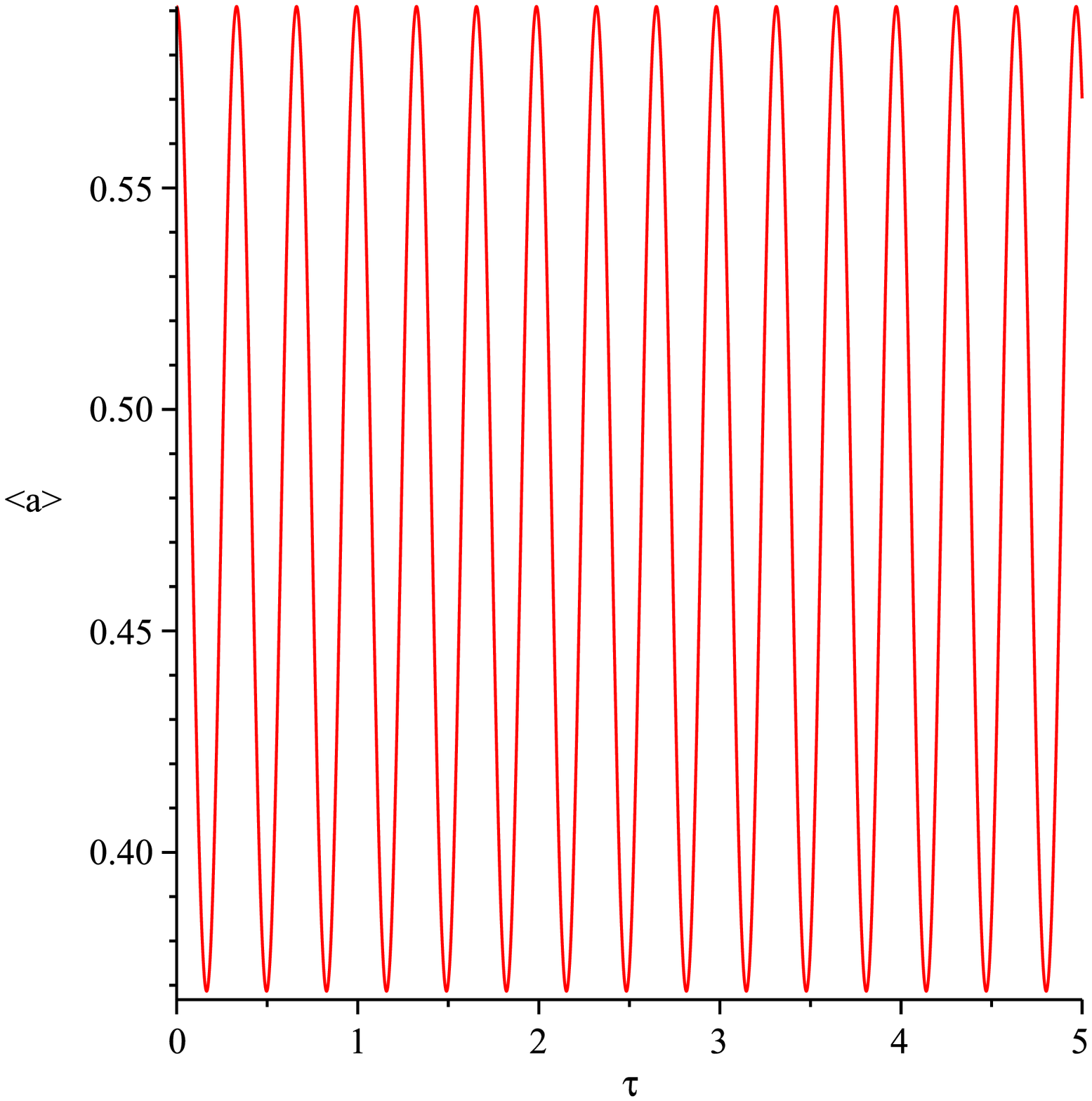}%Gc=90-amedio5
\label{fig2b}
%}
\caption{The expected value of the scale factor $\left<a\right>$. Here, $N=2$, $g_s=-2$, $g_r=120$, $g_\Lambda=0$. We show the cases of $g_{c}=10$ (left) 
and $g_{c}=90$ (right).}
\label{fig2}
\end{figure}

\subsubsection{Behavior of $\left< a \right>$ as $g_\Lambda$ varies}

If we fix $N$ and the HL parameters but $g_\Lambda$, and let $g_\Lambda$ decrease, we observe that:
($i$) the maximum value of $\left<a\right>$ decreases;
($ii$) the amplitude of oscillation of $\left<a\right>$ decreases;
($iii$) the number of oscillations of $\left<a\right>$, for a fixed interval of $\tau$, increases.

That behavior may be understood by the observation of the potential that confines the scale factor. As $g_\Lambda$ decreases, 	
$\left<a\right>$ is forced to oscillate within an ever smaller region. Under those conditions, for fixed $N$ and the other HL parameters, the maximum 
value and the amplitude of $\left<a\right>$ both decrease. Moreover, since the interval where $\left<a\right>$ oscillates is smaller,
the number of oscillations of $\left<a\right>$  for a fixed $\tau$ interval also increases. 
Figs. 3 and 4 illustrate the behavior of the potential $V(a)$ and of the expected value $\left<a\right>$, for two different values of $g_\Lambda$
whereas the $\tau$ interval, $N$ and the other HL parameters remain fixed.

\begin{figure}[h!]
\centering
%\subfloat[$V(a)\times a$ para $g_{c}=10$.]{
\includegraphics[scale=0.3]{%GL=m1V(a)
	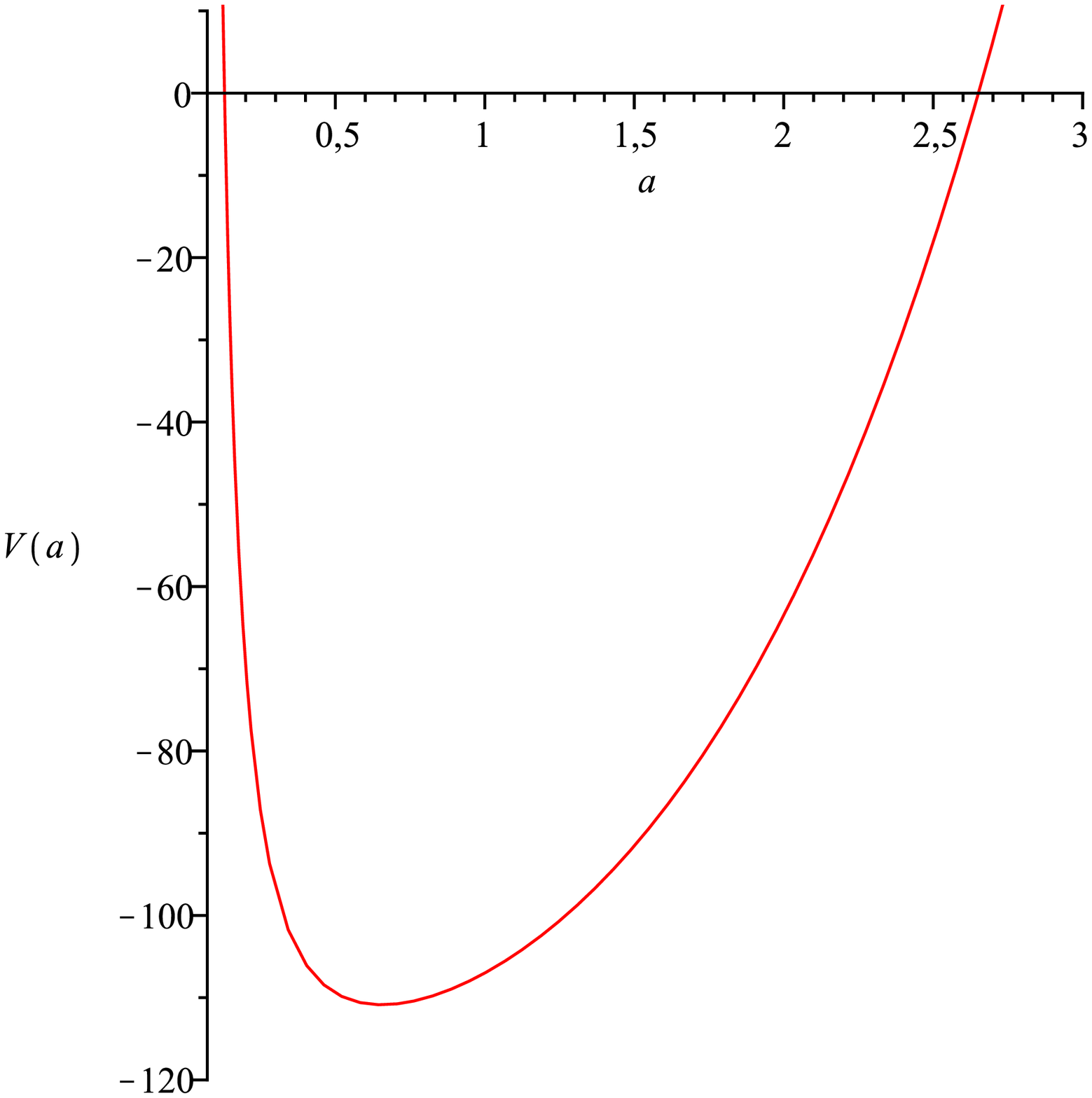}
\label{fig3a}
%}
\quad
%\subfloat[$V(a)\times a$ para $g_{c}=90$.]{
\includegraphics[scale=0.3]{%GL=m25V(a)
	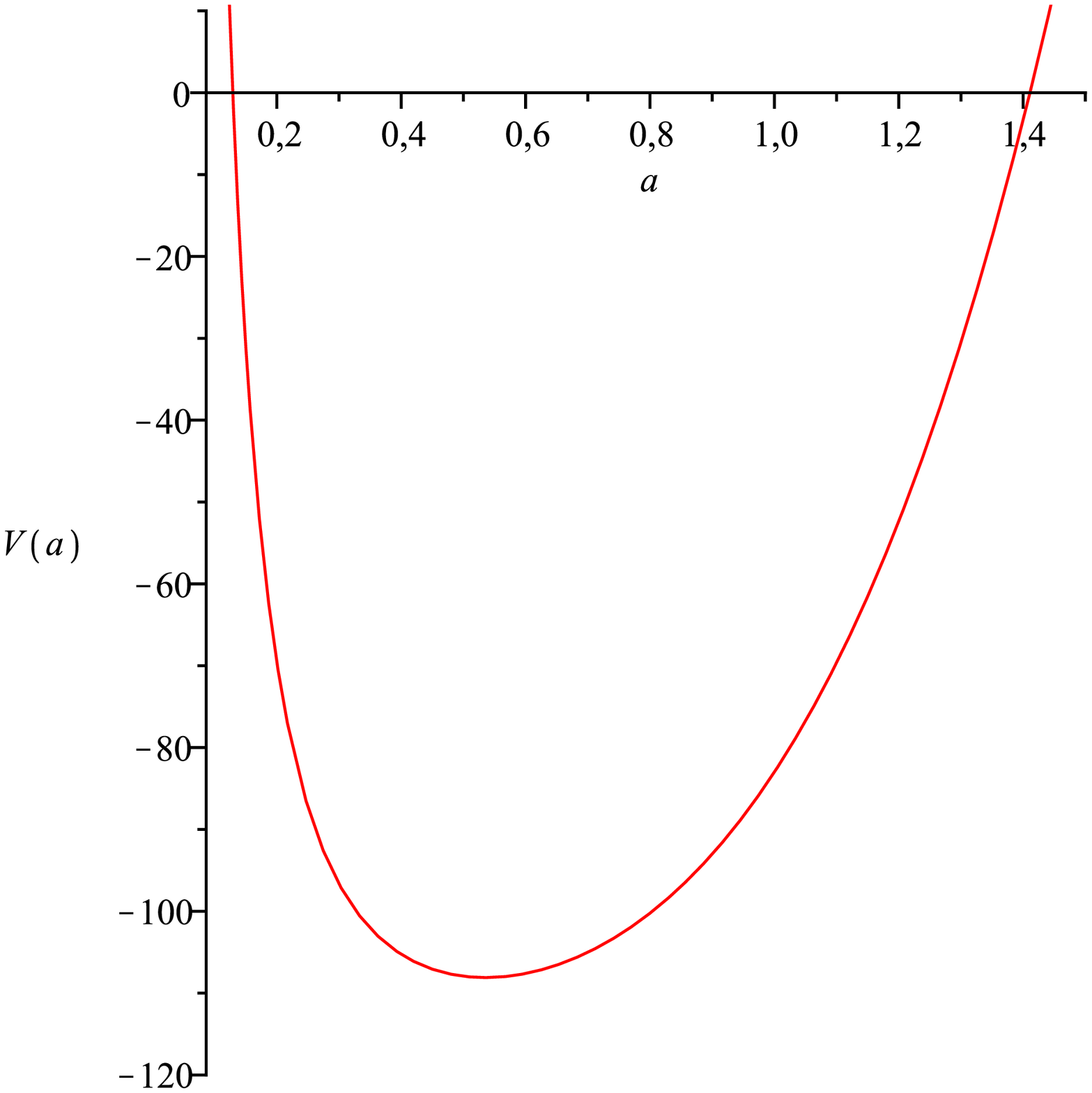}
\label{fig3b}
%}
\caption{The potential $V(a)$ for $g_s=-2$, $g_r=120$, $g_c=10$. We have used $g_\Lambda=-1$ (left) and $g_\Lambda=-25$ (right).}
\label{fig3}
\end{figure}

For $g_\Lambda =-1$, we have $\bar{E}=-103.833980119716$. Therefore, from the potential $V(a)$ (Fig. 3), the interval where $\left<a\right>$ oscillates is\\
$\[0.3676565615,1.137772203\]$, which has an amplitude of 0.7701156415. On the other hand, for $g_\Lambda=-25$, we have $\bar{E}=-95.6220677191430$.
The interval where $\left<a\right>$ oscillates is $\[0.2927449802,0.8679732690\]$, with smaller amplitude, equal to 0.5752282888.
\begin{figure}[h!]
\centering
%\subfloat[$\langle a \rangle\times t$ para $g_{c}=10$.]{
\includegraphics[scale=0.3]{%GL=-1-amedio5
	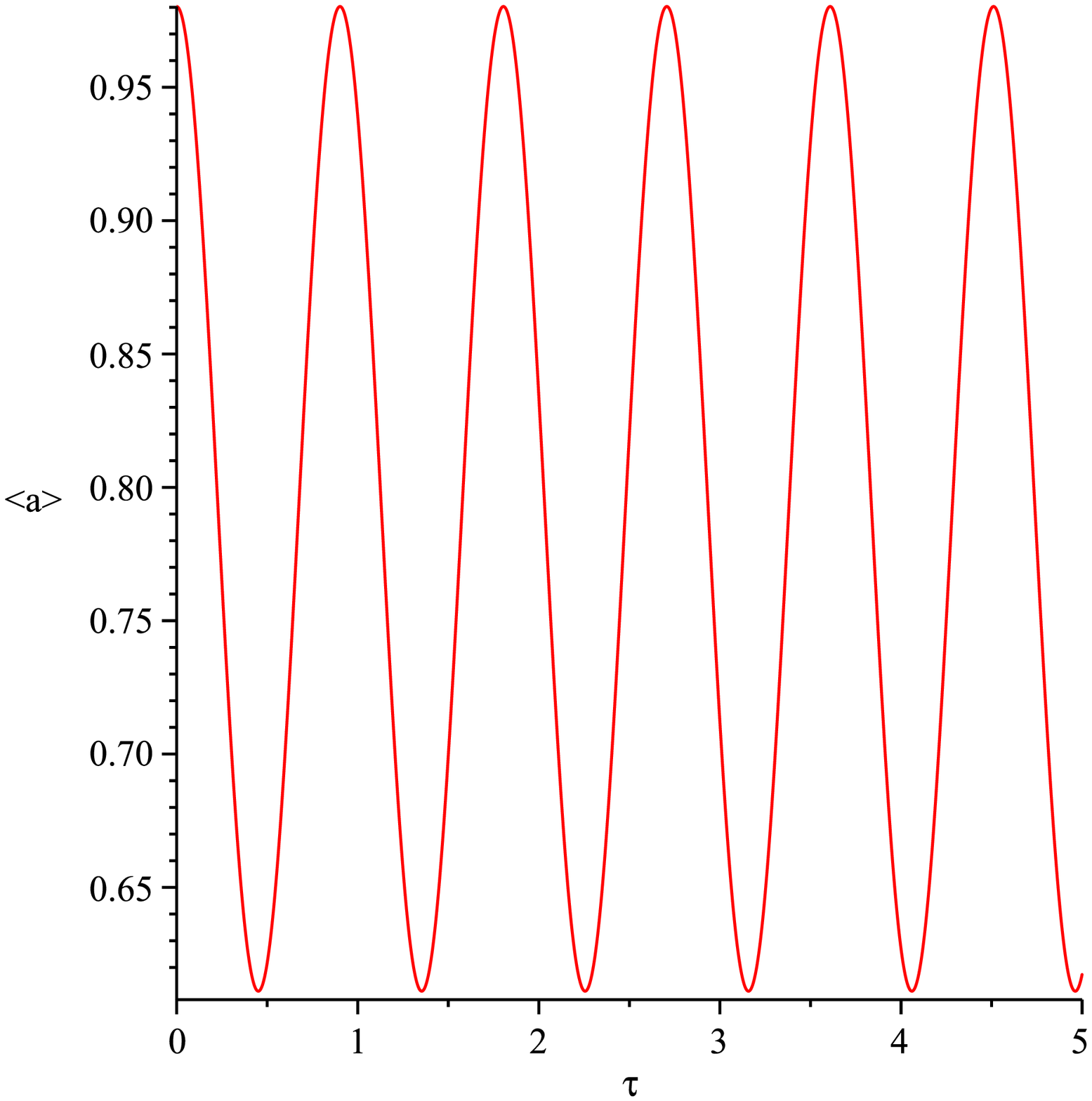}
\label{fig4a}
%}
\quad
%\subfloat[$\langle a \rangle\times t$ para $g_{c}=20$.]{
\includegraphics[scale=0.3]{%GL=-25-amedio5
	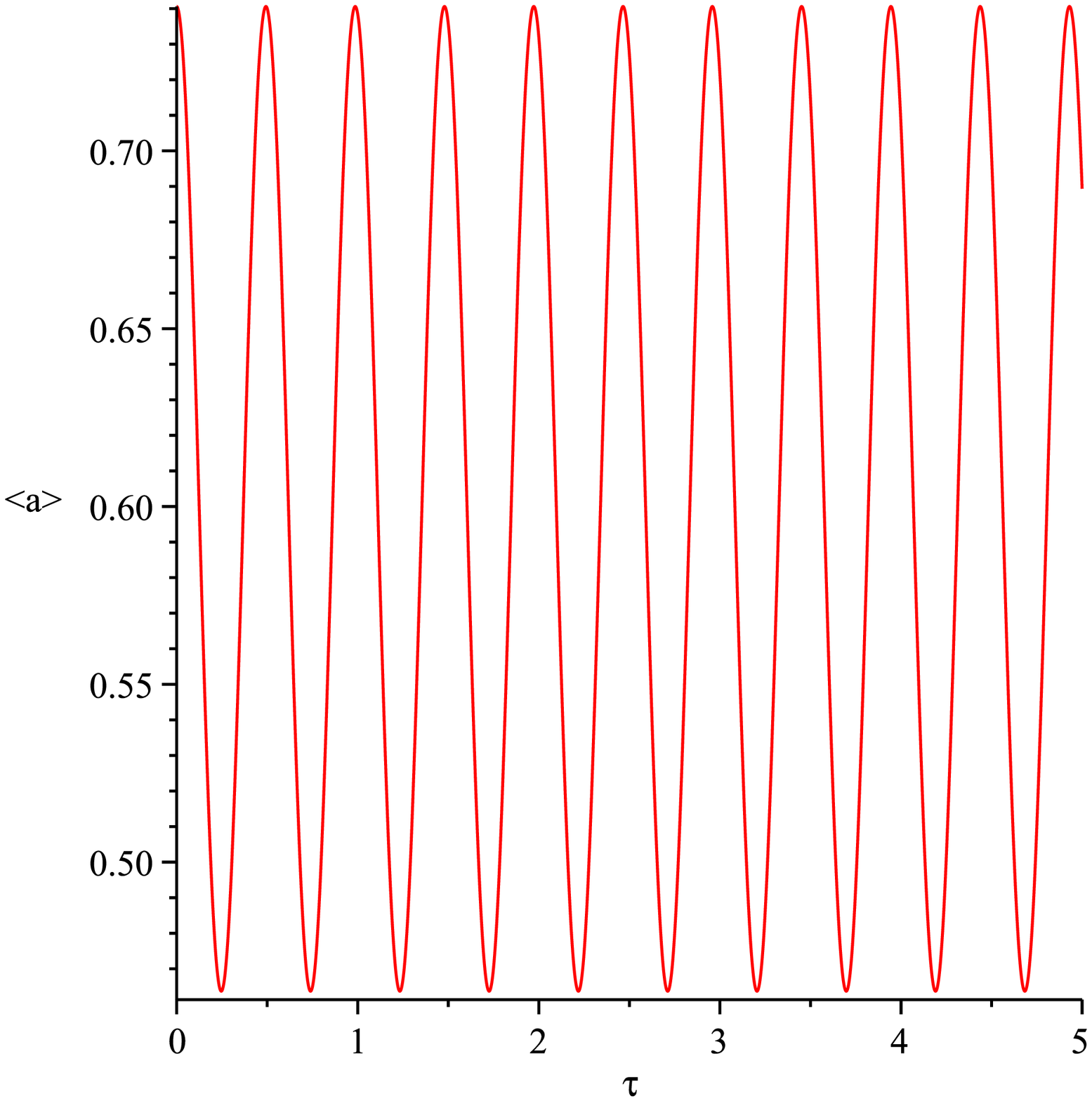}
\label{fig4b}
%}
\caption{The expected value of the scale factor $\left<a\right>$ for $N=2$, $g_s=-2$, $g_r=120$, $g_c=10$. We have used $g_\Lambda=-1$ (left) and $g_\Lambda=-25$ (right).}
\label{fig4}
\end{figure}

\subsubsection{Behavior of $\left< a \right>$ as $g_r$ varies}

If we fix $N$ and the HL parameters but $g_r$, and let $g_r$ vary, we
observe that the maximum value, the amplitude of oscillation and the
number of oscillations of $\left<a\right>$ all remain constant. This
is because as $g_r$ varies, the amplitude of the interval in which
$\left<a\right>$ can oscillate is unaltered. Figs. 5 and 6 show examples
of how $\left<a\right>$ behaves for two different values of $g_r$, whereas
the $\tau$ interval, $N$ and the other HL parameters remain fixed.

\begin{figure}[h!]
\centering
%\subfloat[$V(a)\times a$ para $g_{c}=10$.]{
 \includegraphics[scale=0.3]{%Gr=40V(a)
 	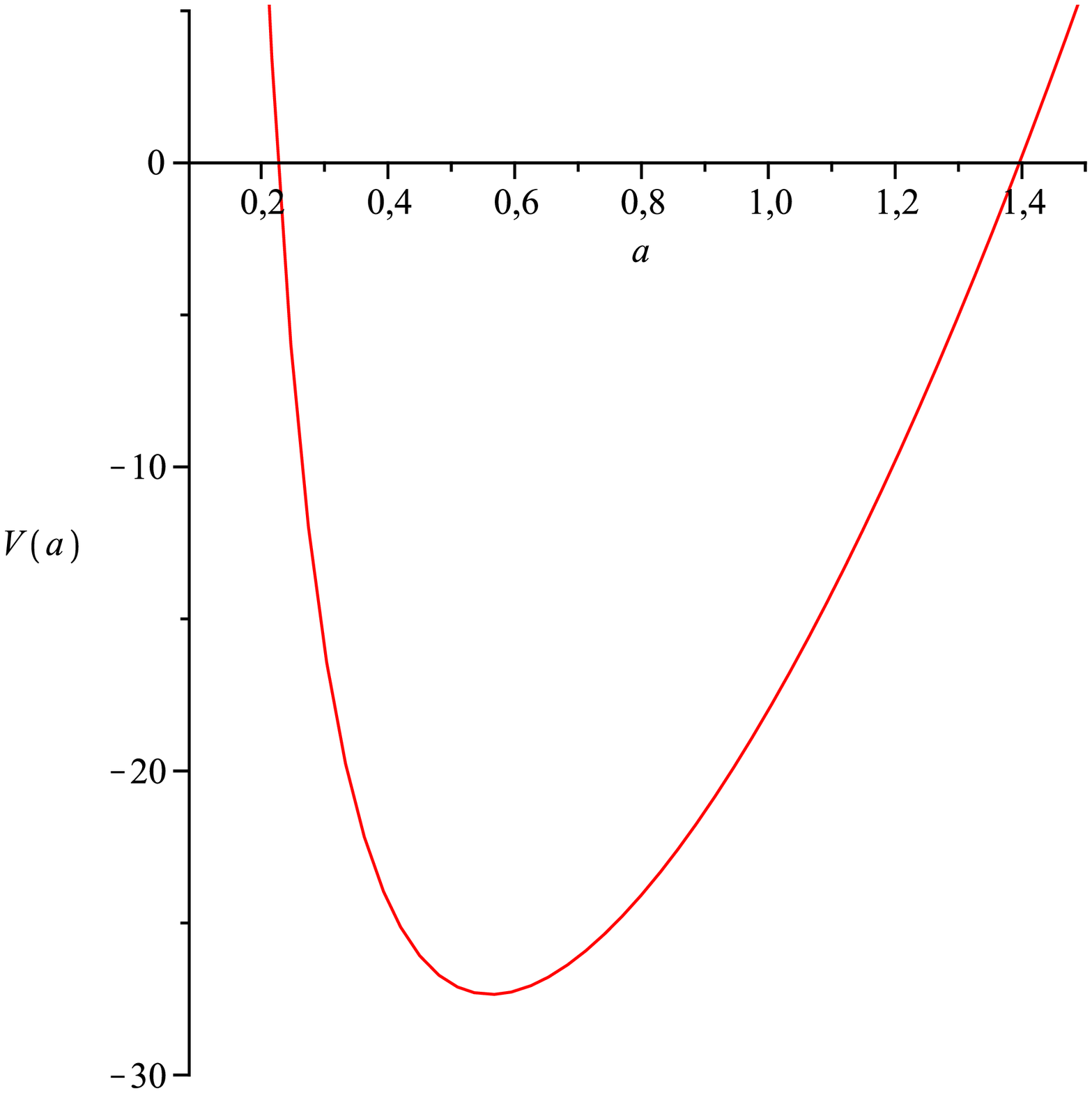}
\label{fig5a}
%}
\quad
%\subfloat[$V(a)\times a$ para $g_{c}=90$.]{
\includegraphics[scale=0.3]{%Gr=80V(a)
	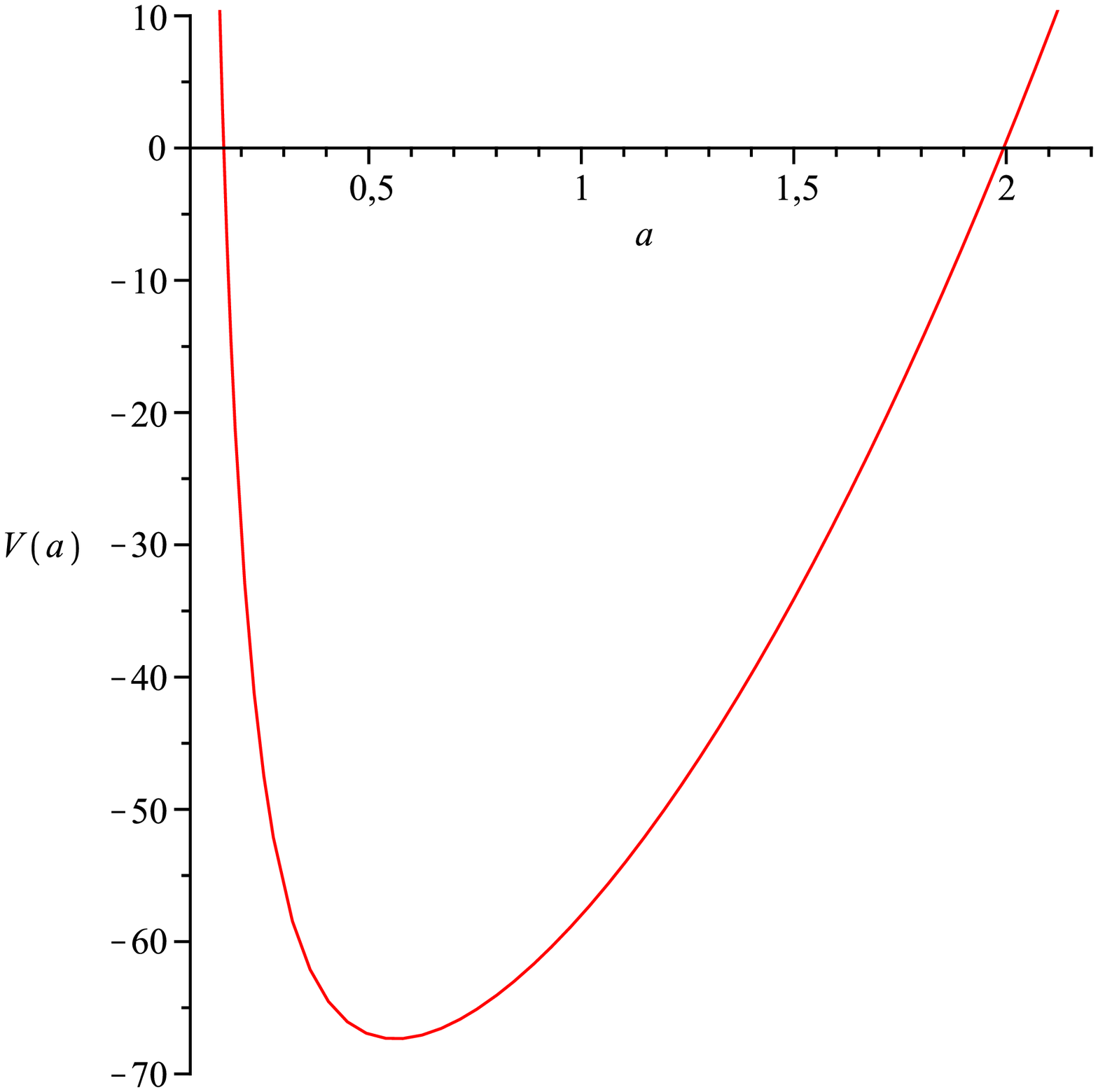}
\label{fig5b}
%}
\caption{The potential $V(a)$ for $g_s=-2$, $g_\Lambda=0$, $g_c=20$. We have used $g_r=40$ (left) and $g_r=80$ (right).}
\label{fig5}
\end{figure}

For $g_r=40$, we have  $\bar{E}=-18.2104751421654$. Therefore, from the potential $V(a)$ (Fig. 5), the interval where $\left<a\right>$ oscillates is\\
$\[0.3180954556,0.9941285249\]$, which has an amplitude of 0.6760330693. On the other hand, for $g_r=80$, we have $\bar{E}=-58.2104751421655$.
The interval where $\left<a\right>$ oscillates is $\[0.3180954556,0.9941285249\]$, which has an amplitude of 0.6760330693.
Identical to the case $g_r=40$.

\begin{figure}[h!]
\centering
%\subfloat[$\langle a \rangle\times t$ para $g_{c}=10$.]{
\includegraphics[scale=0.3]{%Gr=40-amedio5
	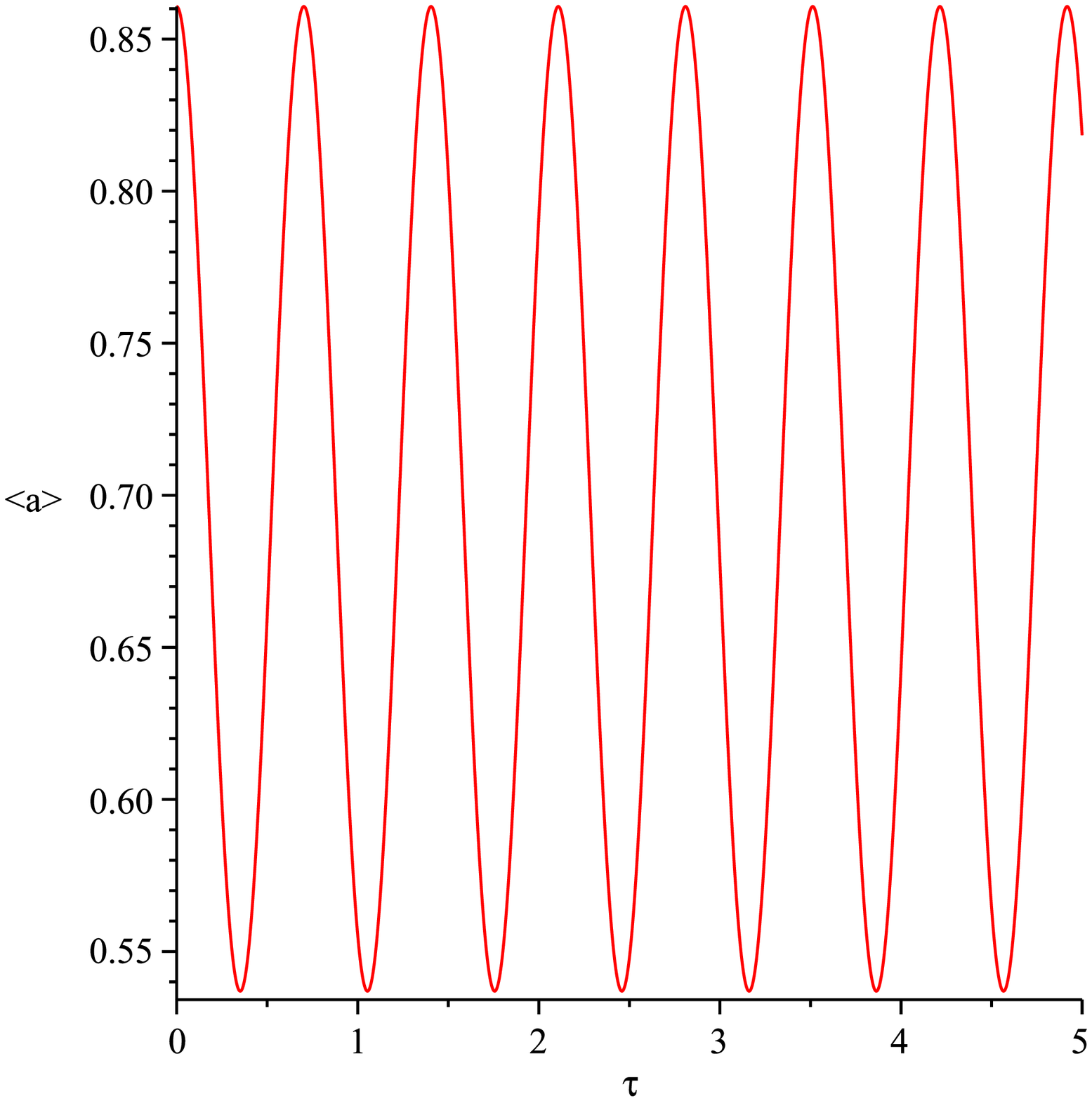}
\label{fig6a}
%}
\quad
%\subfloat[$\langle a \rangle\times t$ para $g_{c}=20$.]{
\includegraphics[scale=0.3]{%Gr=80-amedio5
	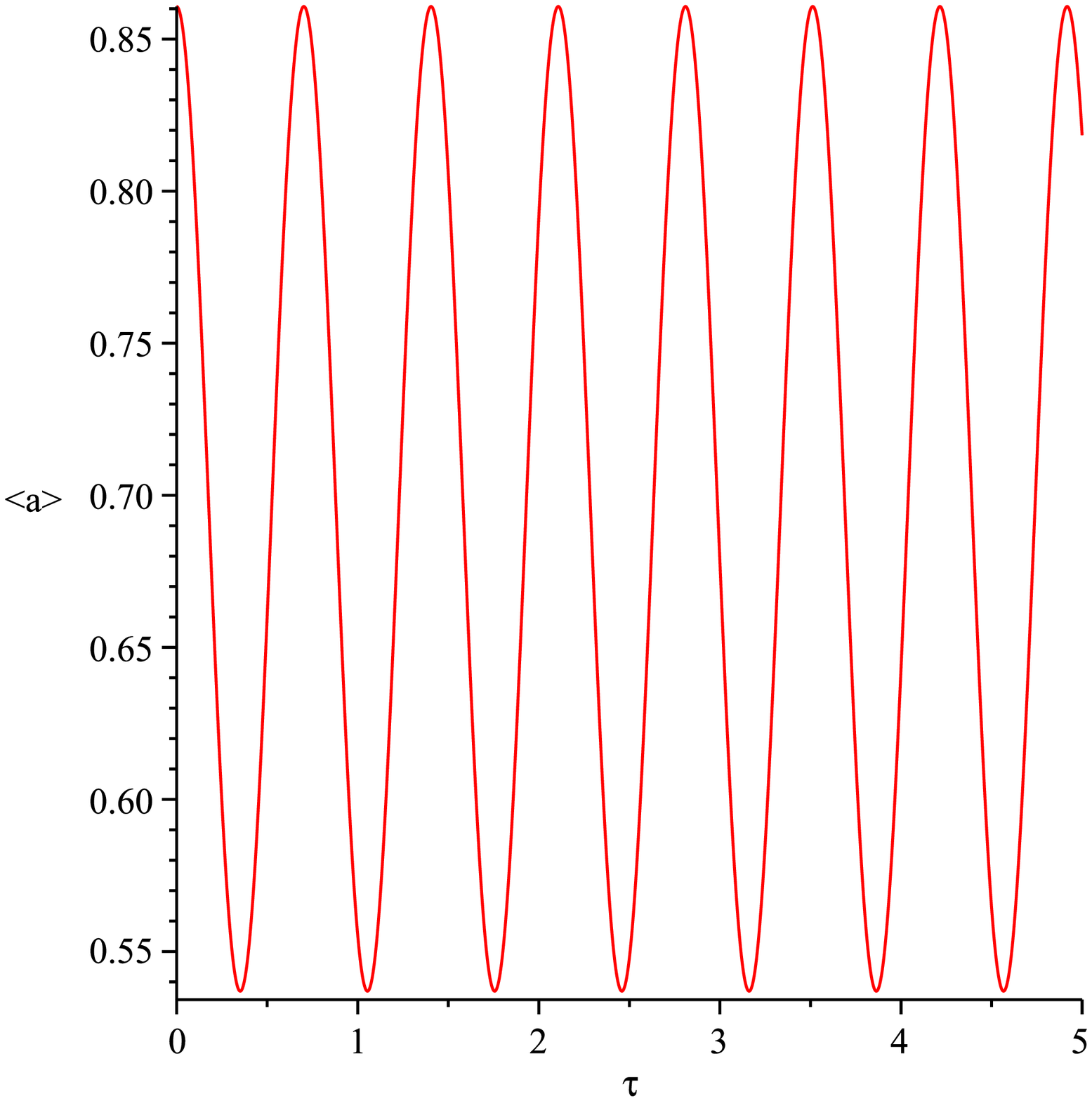}
\label{fig6b}
%}
\caption{The expected value of the scalar factor $\left<a\right>$ for $N=2$, $g_s=-2$, $g_\Lambda=0$, $g_c=20$. We have used $g_r=40$ (left) and $g_r=80$ (right).}
\label{fig6}
\end{figure}

\subsubsection{Behavior of $\left< a \right>$ as $g_s$ varies}

If we fix $N$ and the HL parameters but $g_s$, by letting $g_s$ vary, we observe that:
(i) both the amplitude and the number of oscillation of $\left<a\right>$, for a fixed $\tau$ interval,  remain the same;
(ii) the maximum value of $\left<a\right>$ increases as $g_s$ decreases.

That behavior may be understood by the observation of the potential that confines the scale factor. As $g_s$ decreases, 	
the interval of oscillation of $\left<a\right>$ is unaltered. Under those conditions, for fixed $N$ and the other HL parameters, 
neither the amplitude nor the number of oscillations of $\left<a\right>$ vary. Notwithstanding, decreasing $g_s$
the interval of oscillation of $\left<a\right>$ shifts to the right; hence the maximum value of $\left<a\right>$ gets larger.
Figs. 7 and 8 illustrate the behavior of the potential $V(a)$ and of the expected value $\left<a\right>$, for two different values of $g_s$
whereas the $\tau$ interval, $N$ and the other HL parameters remain fixed.

\begin{figure}[h!]
\centering
%\subfloat[$V(a)\times a$ para $g_{c}=10$.]{
\includegraphics[scale=0.3]{%Gs=m20V(a)
	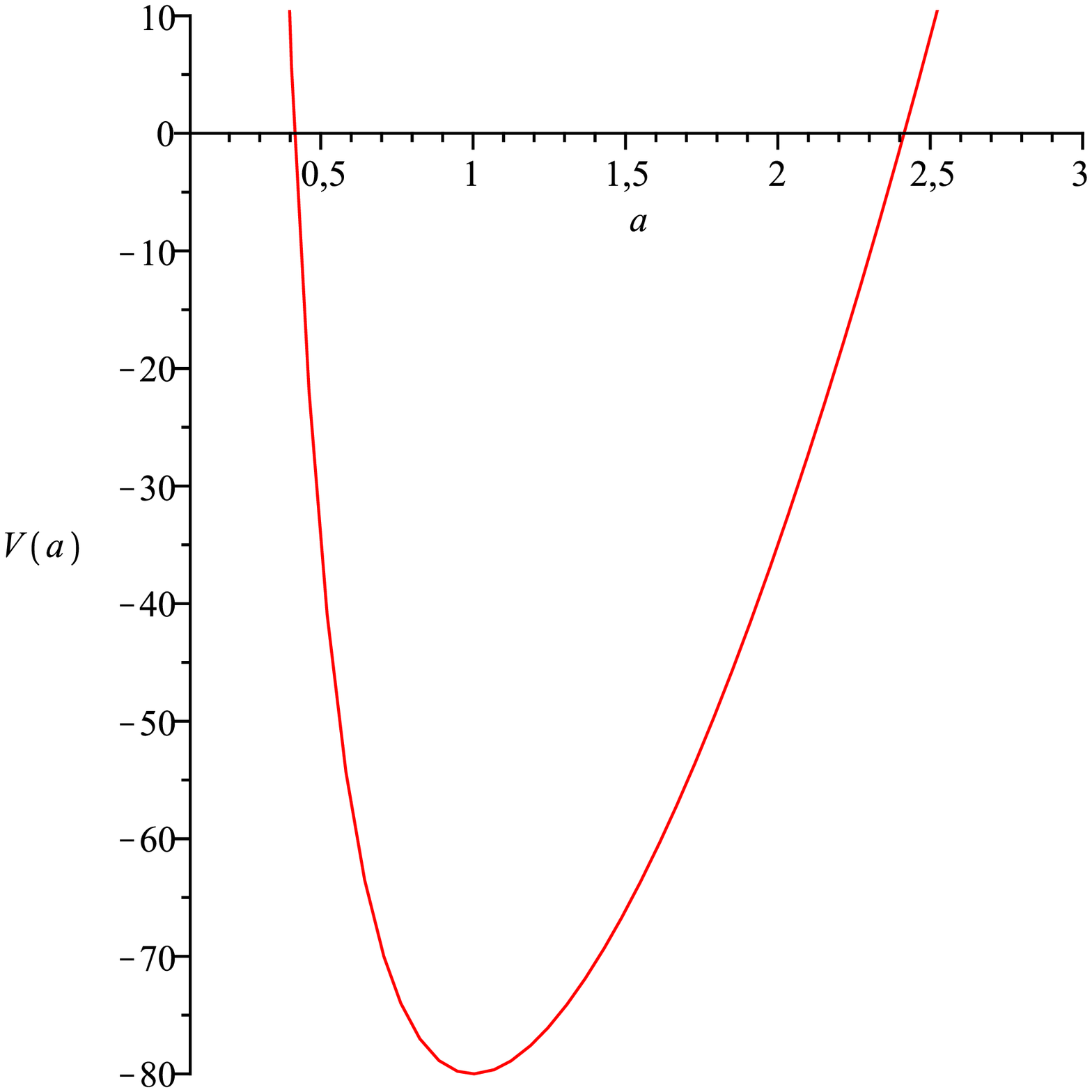}
\label{fig7a}
%}
\quad
%\subfloat[$V(a)\times a$ para $g_{c}=90$.]{
\includegraphics[scale=0.3]{%Gs=m120V(a)
	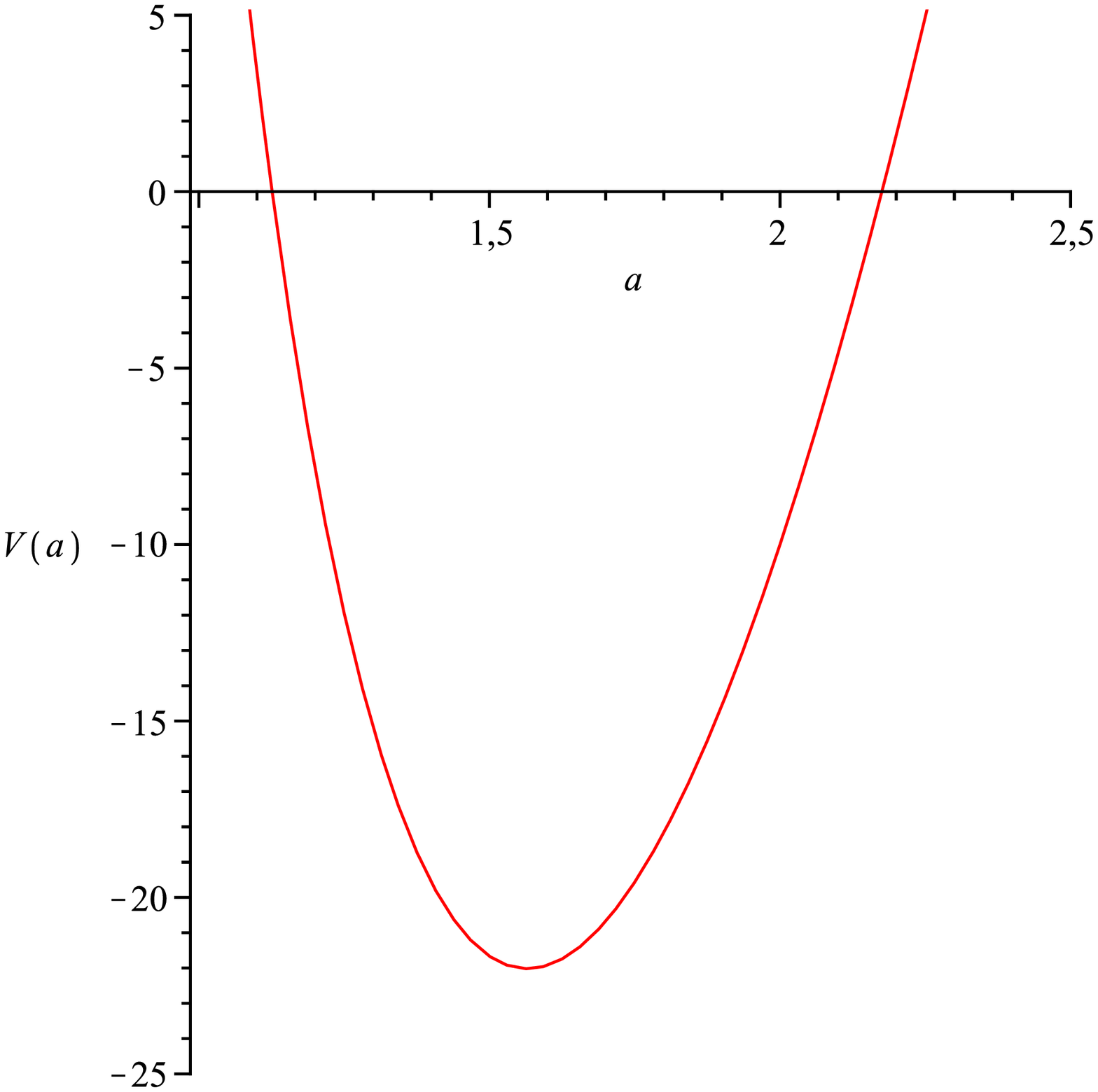}
\label{fig7b}
%}
\caption{The potential $V(a)$ for $g_\Lambda=0$, $g_r=120$, $g_c=20$. We have used $g_s=-20$ (left) and $g_s=-120$ (right).}
\label{fig7}
\end{figure}

For $g_s = -20$, we have $\bar{E}=-70.9932768412420$. Therefore, from the potential $V(a)$ (Fig. 7), the interval where $\left<a\right>$ oscillates is\\
$\[0.7192555414,1.390326445\]$, which has an amplitude of 0.6710709036. On the other hand, for $g_s = -120$, we have $\bar{E}=-13.0506261790150$.
The interval where $\left<a\right>$ oscillates is $\[1.265657001,1.935350368\]$, with amplitude equal to 0.669693367.

\begin{figure}[h!]
\centering
%\subfloat[$\langle a \rangle\times t$ para $g_{c}=10$.]{
\includegraphics[scale=0.3]{%Gs=-20-amedio5
	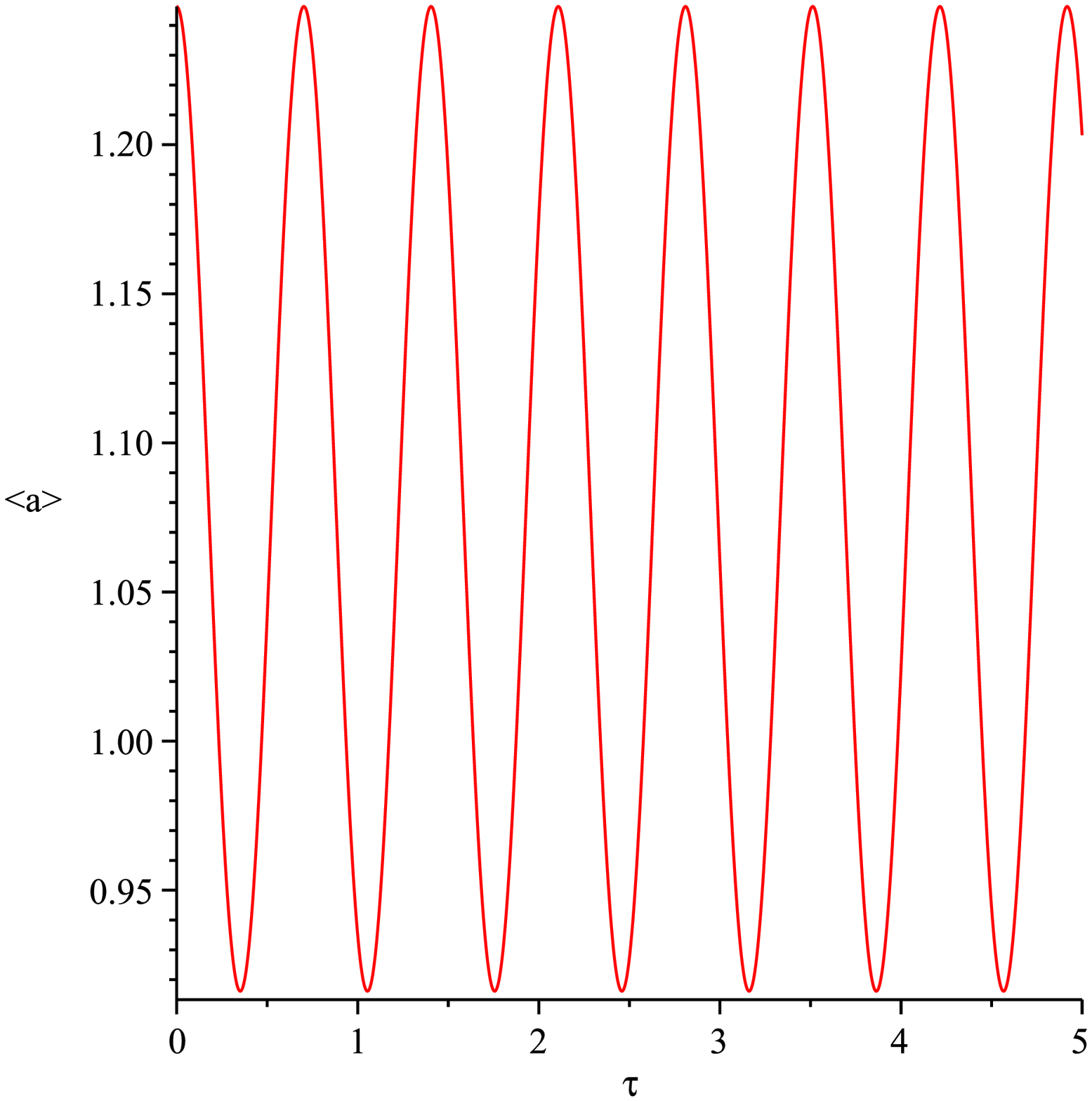}
\label{fig8a}
%}
\quad
%\subfloat[$\langle a \rangle\times t$ para $g_{c}=20$.]{
\includegraphics[scale=0.3]{%Gs=-120-amedio5
	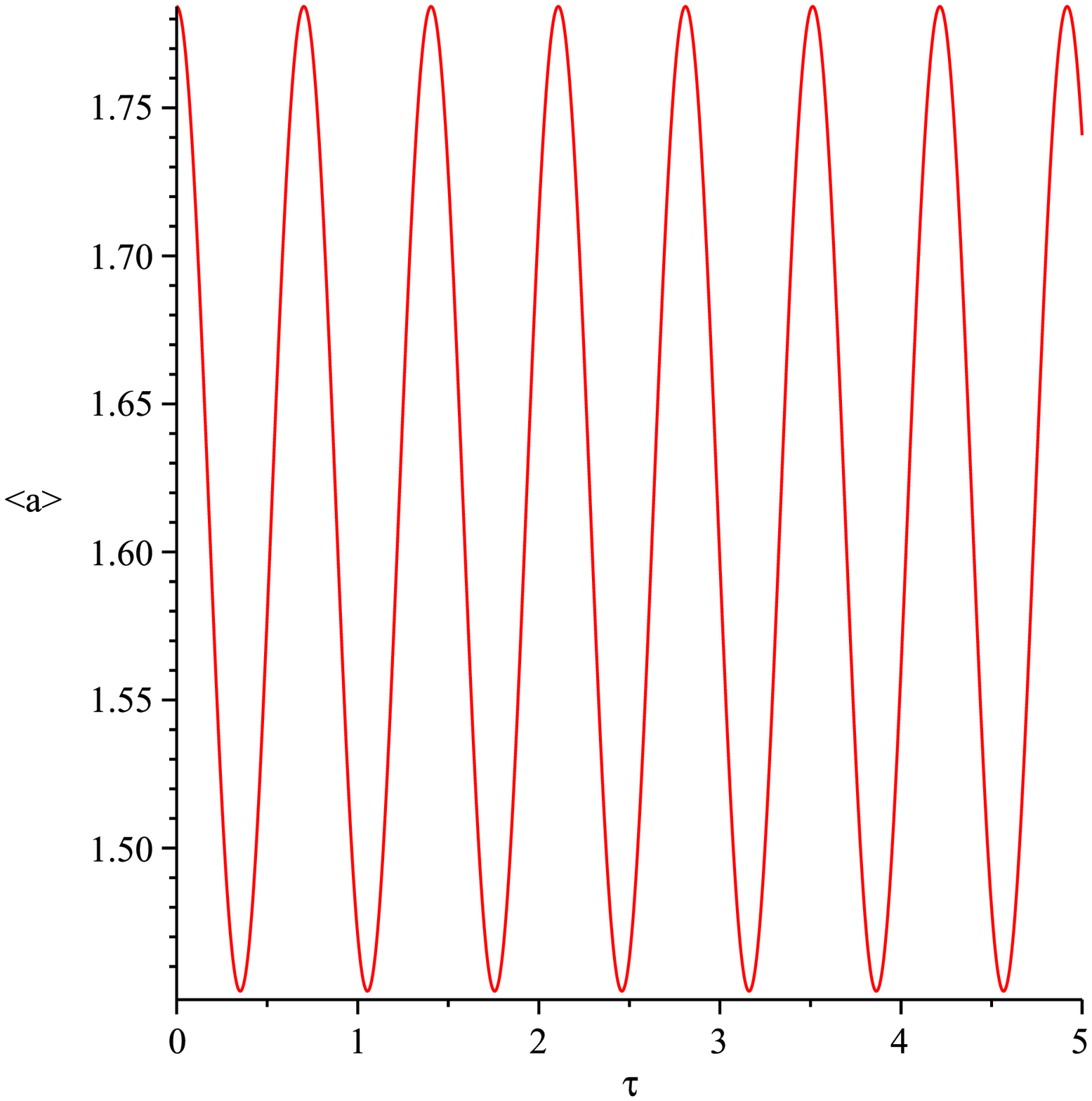}
\label{fig8b}
%}
\caption{The expected value $\left<a\right>$ of the scale factor for $N=2$, $g_\Lambda=0$, $g_r=120$, $g_c=20$. We have used $g_s=-20$ (left) and $g_s=-120$ (right).}
\label{fig8}
\end{figure}

\subsubsection{Behavior of $\left< a \right>$ as $N$ varies}

If we fix all the HL parameters but let $N$ increase, we observe that:
($i$) the maximum value of $\left<a\right>$ increases;
($ii$) the amplitude of oscillation of $\left<a\right>$ increases;
($iii$) the number of oscillations of $\left<a\right>$, for a fixed interval of $\tau$, remains constant.

That behavior may be understood by the observation that the mean energy associated with the wavepacket increases with the increase of $N$. 
As $N$ increases, $\left<a\right>$ oscillates in a ever larger region; hence
its maximum value and the amplitude of the oscillation interval of $\left<a\right>$ both increase. 
The number of oscillations of $\left<a\right>$  for a fixed $\tau$ interval does not change, though. Although the mean energy increases,
the potential energy does not vary; hence only the kinetic energy increases. Therefore, $\left<a\right>$ oscillates
more rapidly in a larger region. The most interesting result is that the oscillation velocity increases in such way that the
oscillation frequency remains constant, as $N$ is increased.

Fig. 1 provides an example for the potential $V(a)$, and Figs. 2 and 9 for $\left<a\right>$, for fixed values of the HL parameters but with different values of $N$.

For $g_c = 10$, with $N=5$, we have $\bar{E}=-95.1047199335060$. Therefore, from the potential $V(a)$ (Fig. 1), the interval where $\left<a\right>$ oscillates is
$\[1.551262998,0.2882899909\]$, which has an amplitude of 1.262973007. On the other hand, for $g_c = 10$ and $N=10$, we have $\bar{E}=-79.2852819937343$.
The interval where $\left<a\right>$ oscillates is $\[2.005428161,0.2230015536\]$, with amplitude, equal to 1.782426607.
Those results must be compared to the results shown in Figs. 1 and 2, in which $g_c = 10$ and $N=2$.
\begin{figure}[h!]
\centering
%\subfloat[$\langle a \rangle\times t$ para $g_{c}=10$.]{
\includegraphics[scale=0.3]{%Gc=10-amedio-N5-L324
	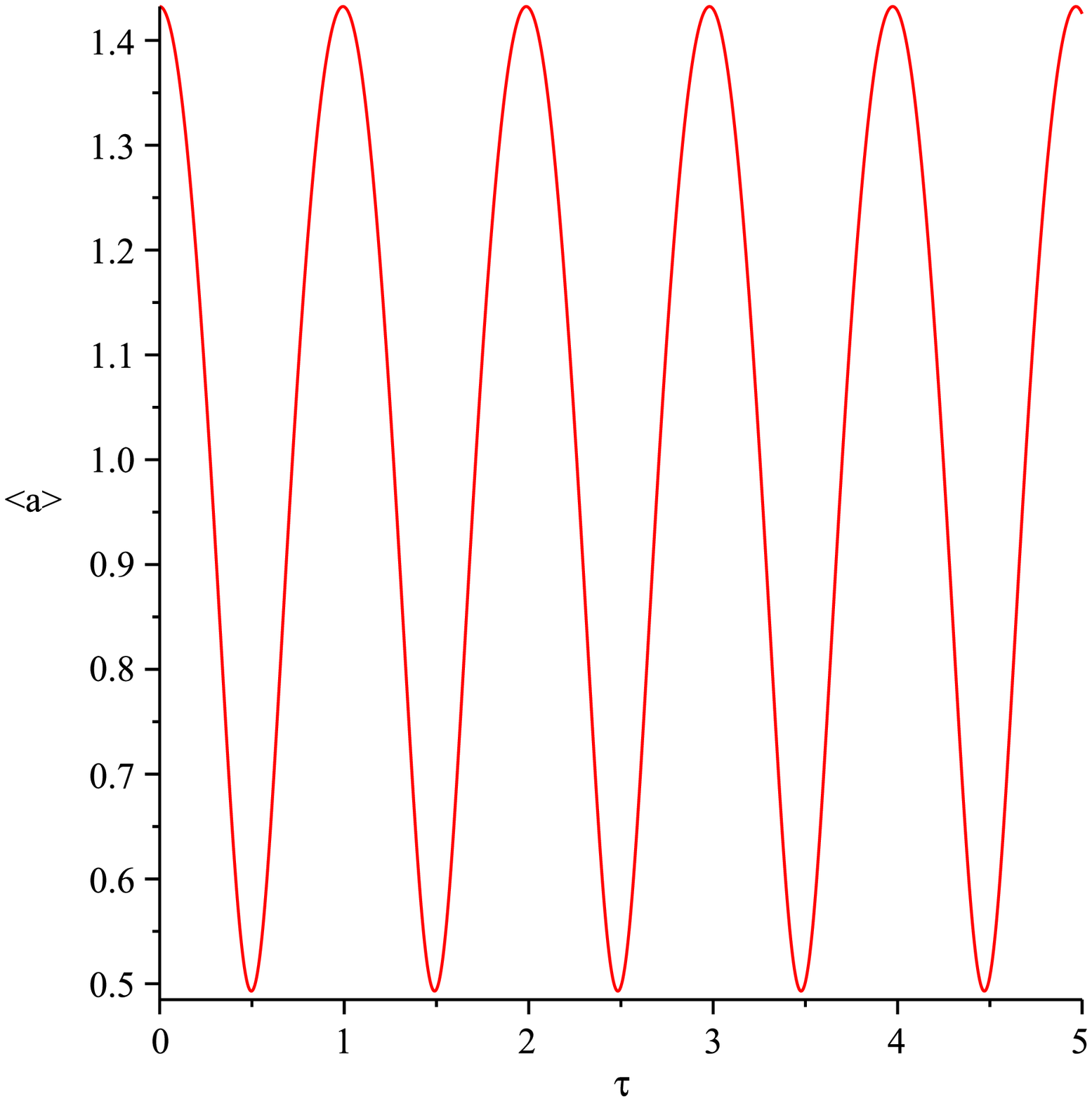}
\label{fig9a}
%}
\quad
%\subfloat[$\langle a \rangle\times t$ para $g_{c}=20$.]{
\includegraphics[scale=0.3]{%Gc=10-amedio-N10-L324
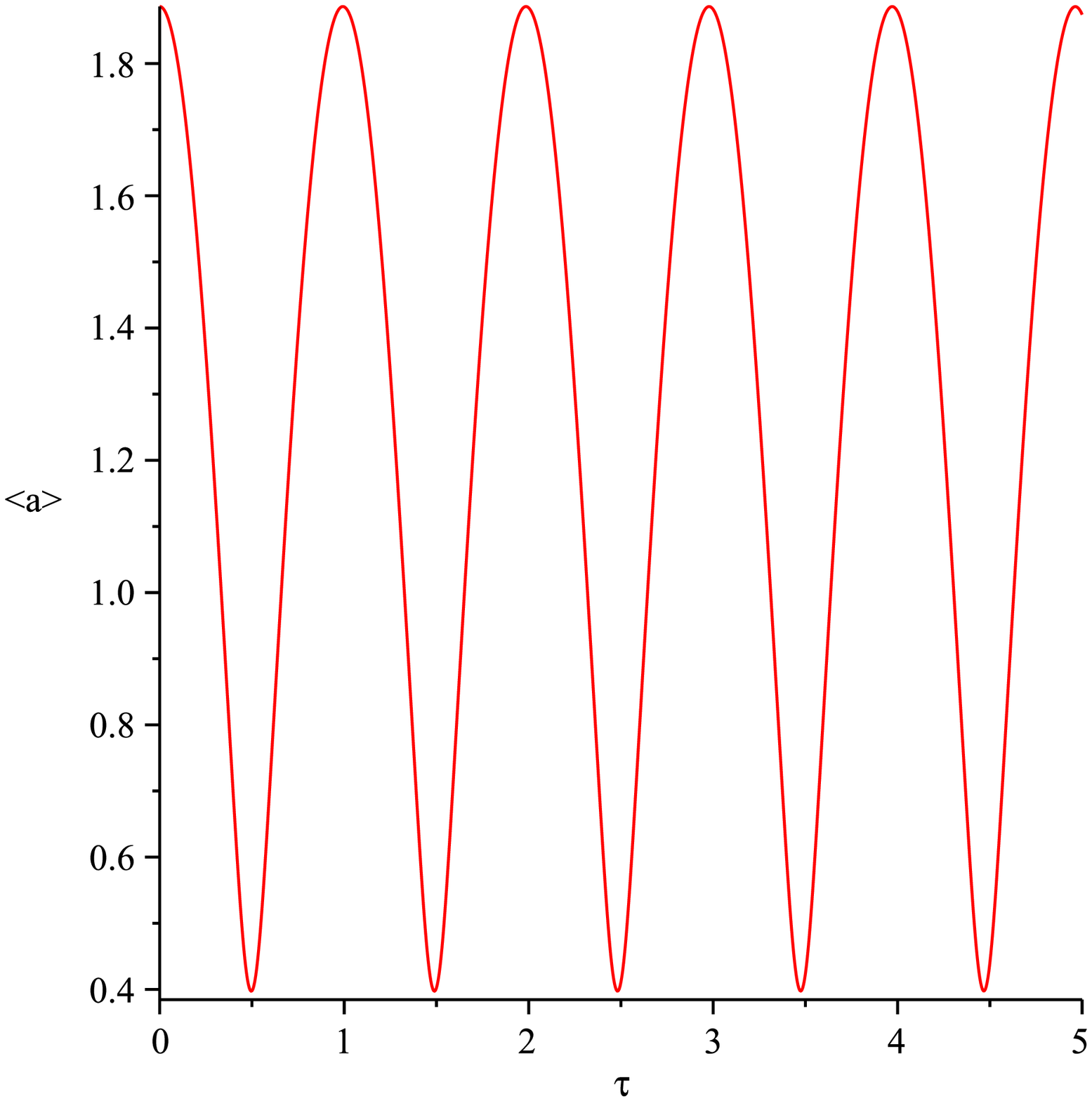}
\label{fig9b}
%}
\caption{The expected value $\left<a\right>$ of the scale factor for $g_c=10$, $g_\Lambda=0$, $g_r=120$, $g_s=-2$. We have used
$N=5$  (left) and $N=10$ (right).}
\label{fig9}
\end{figure}

\subsubsection{Results for the standard deviations}

We calculated $\underline a = \left<a\right>-\sigma_a$, in which
$\left<a\right>$ is given by Eq.(\ref{37}) and $\sigma_a$ by
Eq.(\ref{38}), for different ranges of the HL parameters ($g_c$, $g_r$,
$g_s$, $g_\Lambda$) and values of $N$. For all computed cases,
$\underline a$ is always positive, thus giving a stronger indication that
the present models are free from singularities at the quantum level. As
for the mathematical significance of that result, we should mention that
if our probability distribution were a normal one and if one took the
interval $[\left<a\right> - \sigma_a, \left<a\right> + \sigma_a]$, symmetric about
the mean value $\left<a\right>$, it would cover $68,26\%$ of the area
under the distribution curve \cite{meyer2}. Fig. 10 shows two examples
of how $\left<a\right>$ and $\underline a$ evolve with time.

\begin{figure}[h!]
\centering
%\subfloat[Incerteza e valor esperado para $g_\Lambda=-1$.]{
\includegraphics[scale=0.3]{%Gc=90-incerteza500000-2niveis1000pontos
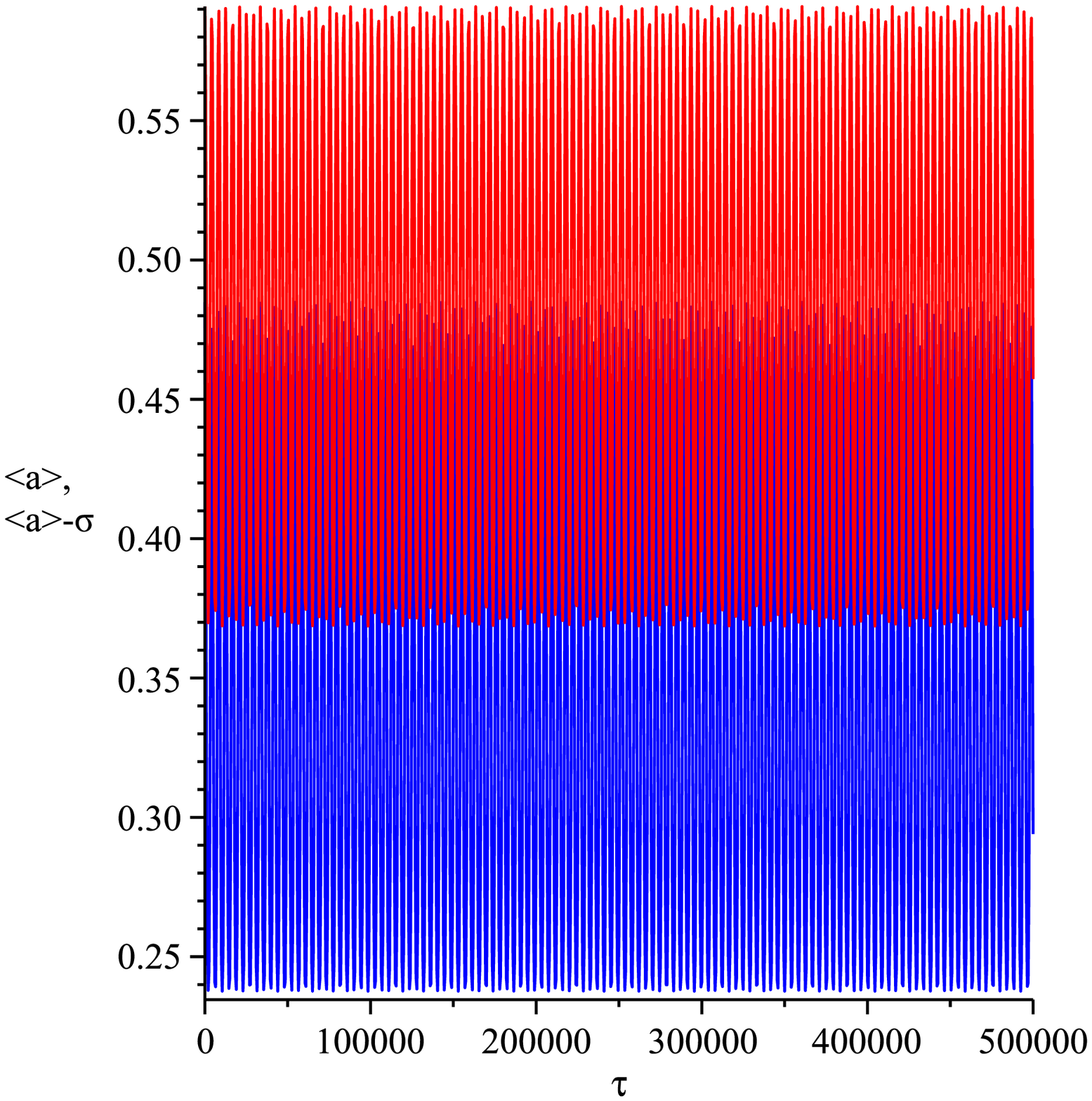}
\label{fig10a}
%}
%\quad
%\subfloat[Incerteza e valor esperado para $g_{c}=10$.]{
\includegraphics[scale=0.3]{%Gs=-20-incerteza500000-10niveis1000pontos
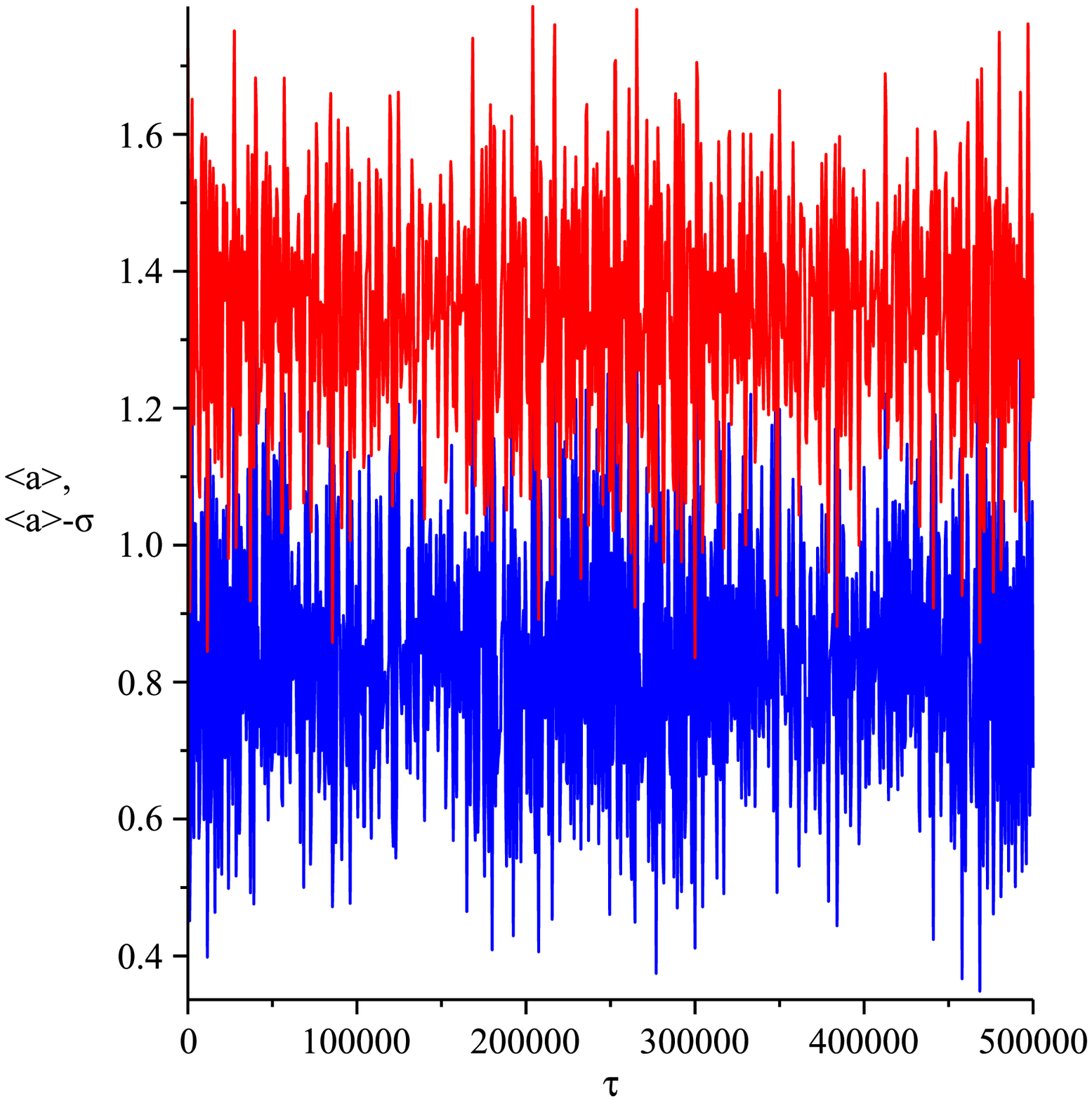}
\label{fig10b}
%}
\caption{The functions $\left<a\right>$ and $\underline a  = \left<a\right> - \sigma_a$.
during the time interval $0<\tau<500000$. Left: we have used $g_c=90$, $g_r=120$, $g_\Lambda=0$, $g_s=-2$ and $N=2$. Right: we have used $g_s=-20$, $g_r=120$, $g_\Lambda=0$, $g_c=20$ and $N=10$.}
\label{fig10}
\end{figure}

\section{DeBroglie-Bohm Interpretation}
\label{debroglie-bohm}

In this section we approach the HL quantum cosmological model using the
{\it DeBroglie-Bohm} interpretation of quantum
mechanics \cite{bohm,holland}. There are several works in quantum
cosmology that have employed such 
interpretation \cite{gil3,kim,acacio1,pedram1,gil,vakili,das,das1,gil2}.
We aim at comparing results obtained from this interpretation with that of
the {\it Many Worlds} interpretation, mainly that the model is free from
singularities at the quantum level.

The first step of applying DeBroglie-Bohm interpretation is that of
rewriting the quantum cosmological wavefunction $\Psi (a,\tau)$ in its
polar form \cite{holland} \begin{equation}
\Psi(a,\tau)=R(a,\tau) e^{iS(a,\tau)},
\label{40}
\end{equation}
in which $R(a,\tau)$ and $S(a,\tau)$ are the amplitude and phase of the
wavefunction, respectively. Equations for $R(a,\tau)$ and $S(a,\tau)$
can be obtained by inserting $\Psi (a,\tau)$ of Eq. (\ref{40}) into Eq.
(\ref{21}).
Following Refs. \cite{holland} and \cite{kim1}, we then obtain two independent equations: one from the real part,
\begin{equation}
\label{41}
 \left(\frac{\partial S(a,\tau)}{\partial a}\right)^{2}+Q(a,\tau)+4\left(g_{c}ka^{2}-g_{\Lambda}a^{4}-g_{r}k^{2}-\frac{g_{s}k}{a^{2}}\right) = 0,
\end{equation}
and another from the imaginary part,
\begin{equation}
\label{41a}
2\frac{\partial{R(a,\tau)}}{\partial a}\frac{\partial S(a,\tau)}{\partial a} + R(a,\tau)\frac{\partial^{2} S(a,\tau)}{\partial a^{2}}+\frac{4}{a^{3\omega-1}}\left(\frac{\partial R(a,\tau)}{\partial \tau}\right)=0.
\end{equation}

In Eq. (\ref{41}), the function $Q(a,\tau)$ is known as the
{\it Bohmian quantum potential}. From the calculations leading to Eq. (\ref{41}) one obtains that
$Q(a,\tau)$ is given by,
\begin{equation}
\label{42}
Q(a,\tau)=-\frac{1}{R(a,\tau)}\frac{\partial^2 R(a,\tau)}{\partial a^2}.
\end{equation}
The {\it Bohmian trajectory} of the scale factor $a$ is \cite{holland}
\begin{equation}
\label{43}
\frac{da(\tau)}{d\tau}=\frac{1}{M}\frac{\partial S}{\partial a},
\end{equation}
in which, from Eq. (\ref{16}), we have the mass $M=2$.

In DeBroglie-Bohm interpretation, the quantum behavior of the Universe is described by the solution of Eq. (\ref{43}). 
To each given initial value of $a(\tau)$ corresponds a deterministic scale factor trajectory, representing the evolution of the
Universe at the Planck scale.

\subsection{Bohmian trajectories of the scale fator $a$}

Using the wavepacket determined in Eq. (\ref{26}), we have obtained its polar form, Eq. (\ref{40}), identifying its amplitude and phase. 
Inserting its phase $S(a,\tau)$ in Eq. (\ref{43}), we computed the Bohmian trajectories of $a$,
for different values of all HL parameters. We have used here the same procedure as that of the previous section, in order to investigate how the Bohmian trajectories 
of $a$ depend on the HL parameters. We fixed all parameters but one, and let that
parameter vary over a wide range of values. Then, we repeated the calculation,  in the same manner, for all HL parameters. 
Eq. (\ref{43}) has been solved, then, for many  different values of $g_c$, $g_\Lambda$, $g_r$, $g_s$ and $N$. 
For all values,  the qualitative behavior of the Bohmian trajectories of $a(\tau)$ were the same.
They oscillate between maxima and minima values and never vanished. Therefore, in the same way as in the {\it Many Worlds} interpretation,
as we saw in the previous section, in the {\it DeBroglie-Bohm} interpretation those models are free from singularities. 
We have also noticed that the Bohmian trajectories of $a$ are, qualitatively, very similar 
to the corresponding expected values of $a$. That result helps verifying the equivalence between
both quantum mechanical interpretations. 

In what follows, we compare some Bohmian trajectories of $a$ with their
corresponding expected values of $a$. In order to better compare those
two quantum mechanical interpretations we used, for each model, as initial conditions
for $a(\tau)$ at $\tau=0$, in the Bohmian trajectories of $a$, the
expected values of $a$ at $\tau=0$.

\subsubsection{Bohmian trajectories of $a$ as $g_c$ varies}

Solving equation (\ref{43}) for several different values of $g_c$, $N$
and various $\tau$ intervals, while keeping fixed the other HL
parameters, we have observed the following properties of the Bohmian
trajectories of $a$, as $g_c$ increases: ($i$) the maximum value of $a$
decreases; ($ii$) the amplitude of oscillation of $a$ decreases; ($iii$)
the number of oscillations of $a$, for a fixed $\tau$ interval,
increases. Such behavior, exemplified in Figs. 2 and 11, agrees with that
of the expected value $\left<a\right>$, described in the previous
section.

\begin{figure}[h!]
\centering
%\subfloat[Incerteza e valor esperado para $g_\Lambda=-1$.]{
\includegraphics[scale=0.3]{%Gc=10-axt
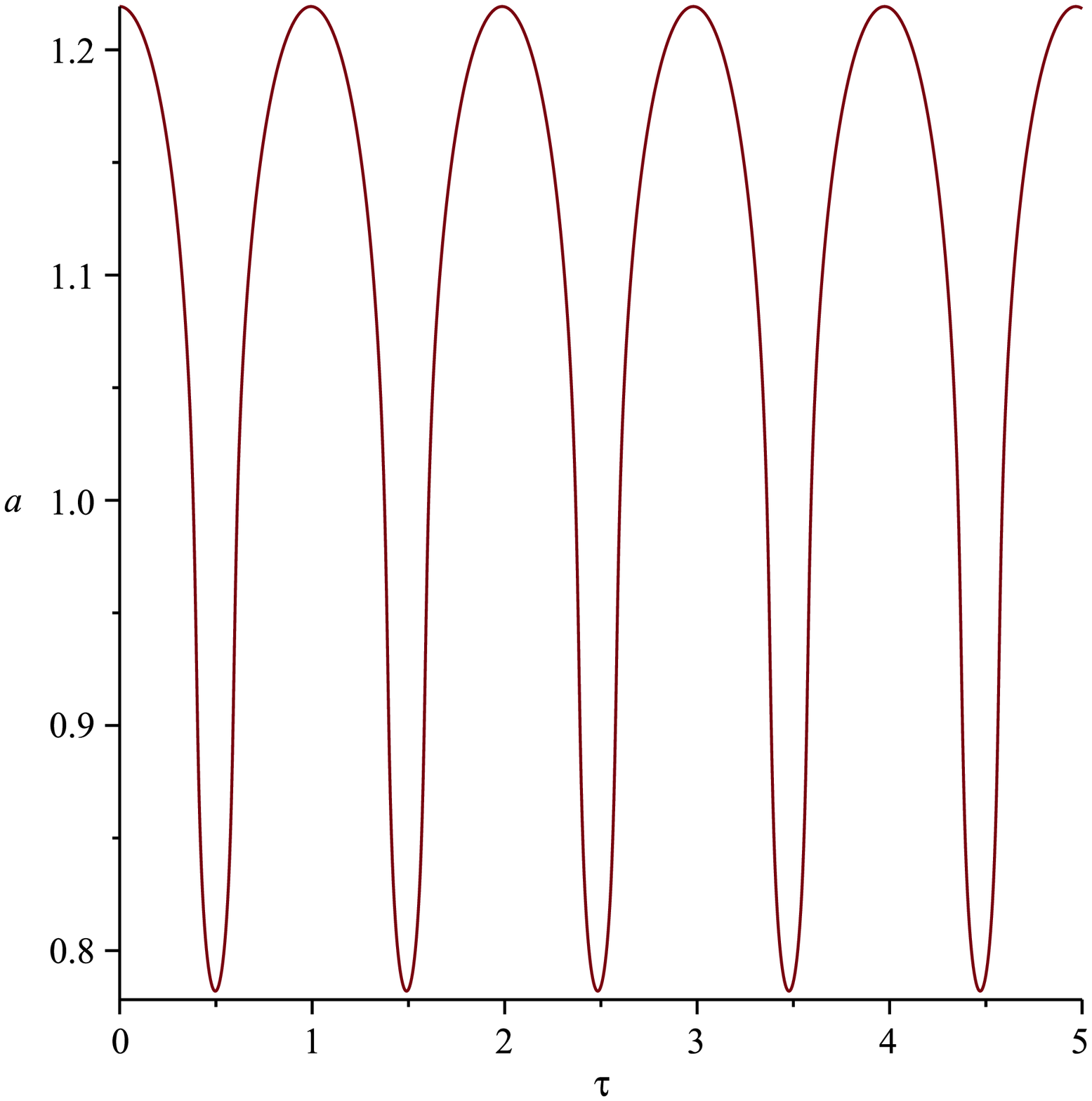}
\label{fig11a}
%}
%\quad
%\subfloat[Incerteza e valor esperado para $g_{c}=10$.]{
\includegraphics[scale=0.3]{%Gc=90-axt
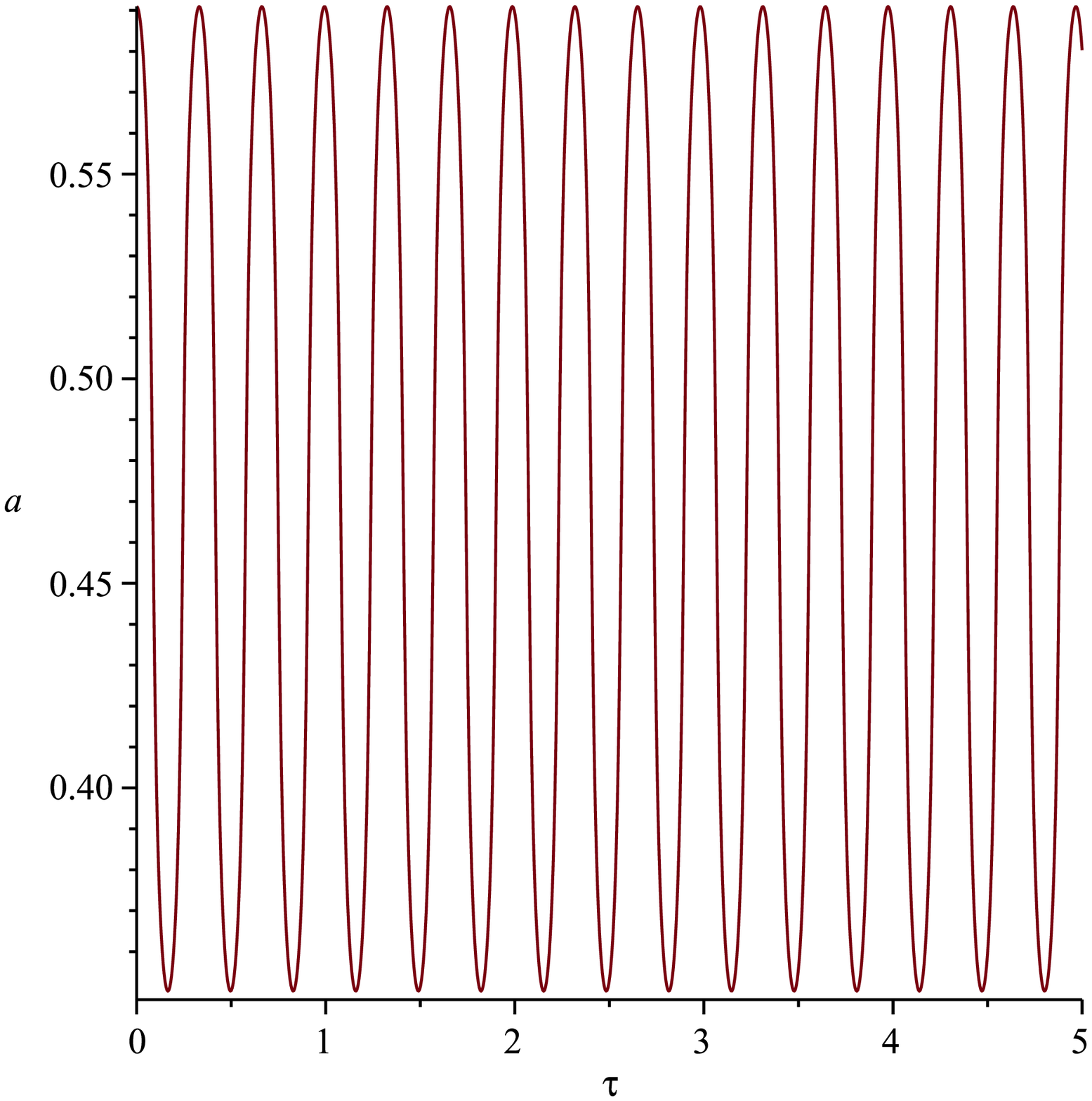}
\label{fig11b}
%}
\caption{Bohmian trajectories of $a$ for $N=2$, $g_s=-2$, $g_r=120$, $g_\Lambda=0$. Left: we have $g_c=10$. Right: we have $g_c=90$.}
\label{fig11}
\end{figure}

\subsubsection{Bohmian trajectories of $a$ as $g_\Lambda$ varies}

Solving equation (\ref{43}) for several different values of $g_\Lambda$, $N$ and various $\tau$ intervals, while keeping fixed the other HL parameters, 
we have observed the following behavior of the Bohmian trajectories of $a$, as
$g_\Lambda$ decreases: 
($i$) the maximum value of $a$ decreases;
($ii$) the amplitude of oscillation of $a$ decreases;
($iii$) the number of $a$ oscillations, for a fixed $\tau$ interval, increases. 
Such behavior of the Bohmian trajectories agrees with that of the expected value $\left<a\right>$, obtained in the previous section. 
Figures 4 and 12 exemplify such agreement.

\begin{figure}[h!]
\centering
%\subfloat[Incerteza e valor esperado para $g_\Lambda=-1$.]{
\includegraphics[scale=0.3]{%GL=-1-axt
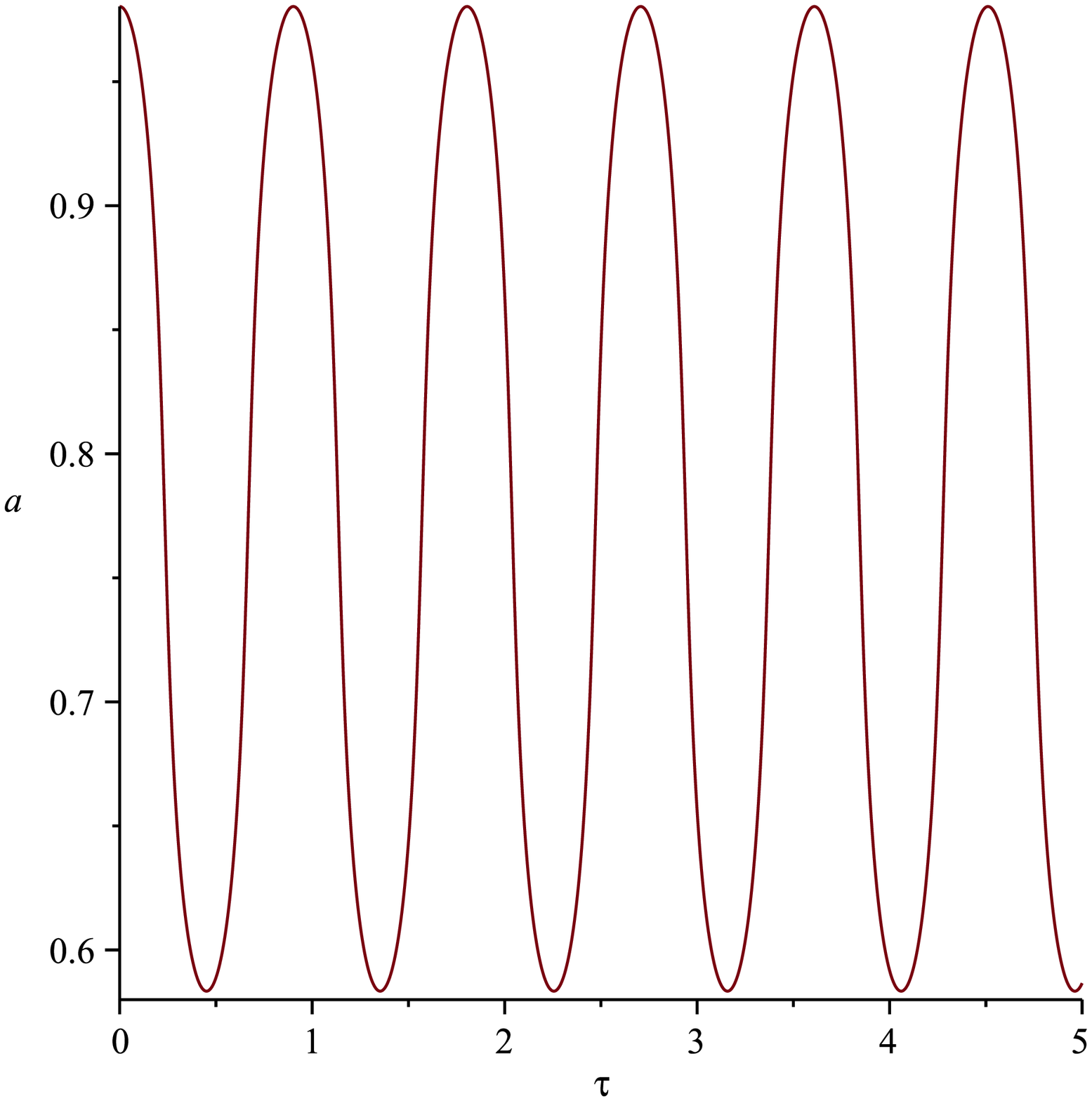}
\label{fig12a}
%}
%\quad
%\subfloat[Incerteza e valor esperado para $g_{c}=10$.]{
\includegraphics[scale=0.3]{%GL=-25-axt
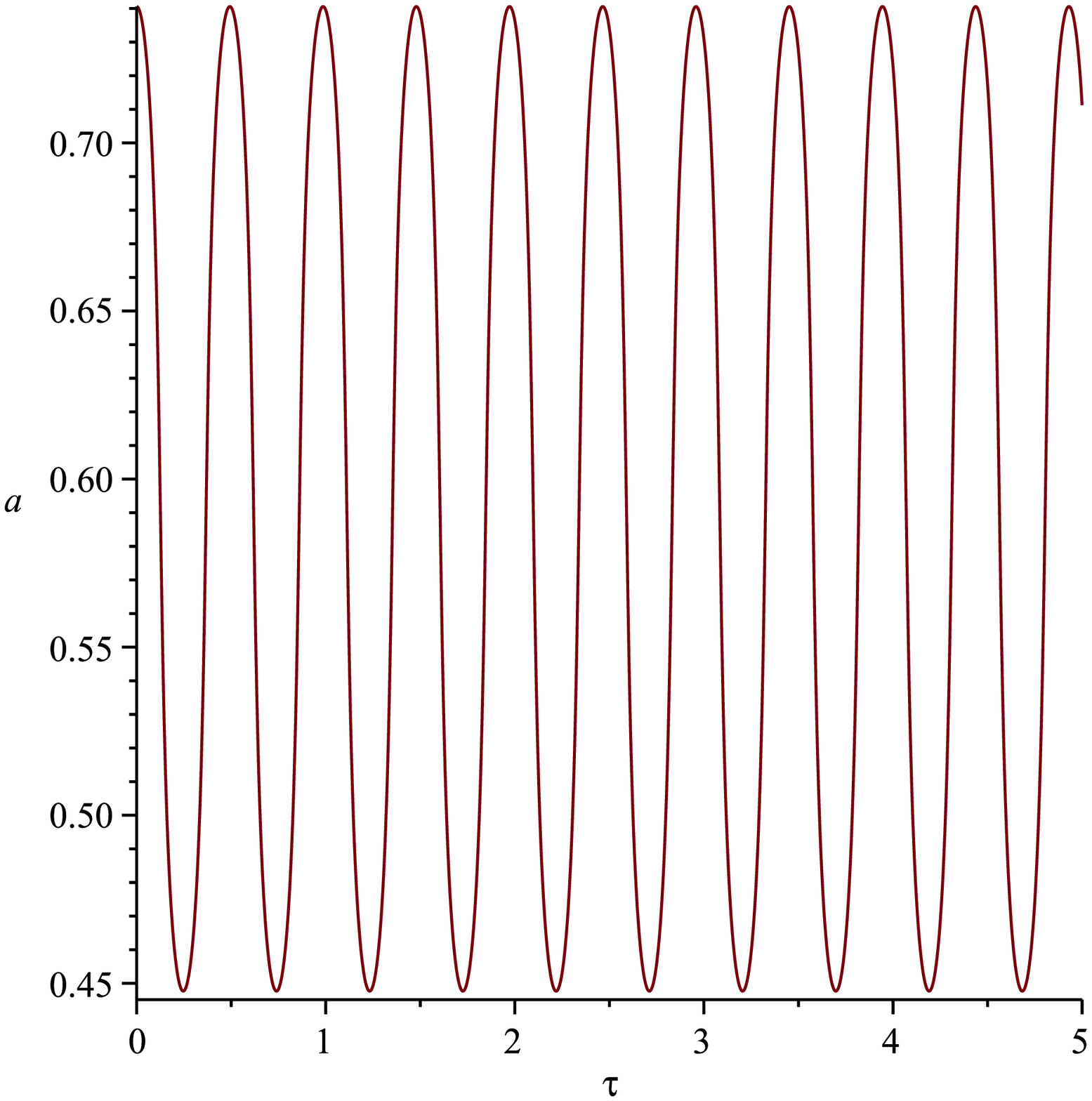}
\label{fig12b}
%}
\caption{Bohmian trajectories of $a$ for $N=2$, $g_s=-2$, $g_r=120$, $g_c=10$. Left: we have $g_\Lambda=-1$. Right: we have $g_\Lambda=-25 $.}
\label{fig12}
\end{figure}

\subsubsection{Bohmian trajectories of $a$ as $g_r$ varies}

Solving equation (\ref{43}) for several different values of $g_r$, $N$ and various $\tau$ intervals, while keeping fixed the other HL parameters, 
we have observed that the maximum value, the amplitude of oscillation and the number of $a$ oscillations do not vary as $g_r$ varies, 
for a fixed $\tau$ interval. Such behavior agrees with that of the expected value of $a$, obtained in the previous section.
Figs. 6 and 13 illustrate that agreement. 

\begin{figure}[h!]
\centering
%\subfloat[Incerteza e valor esperado para $g_\Lambda=-1$.]{
\includegraphics[scale=0.3]{%Gr=40-axt
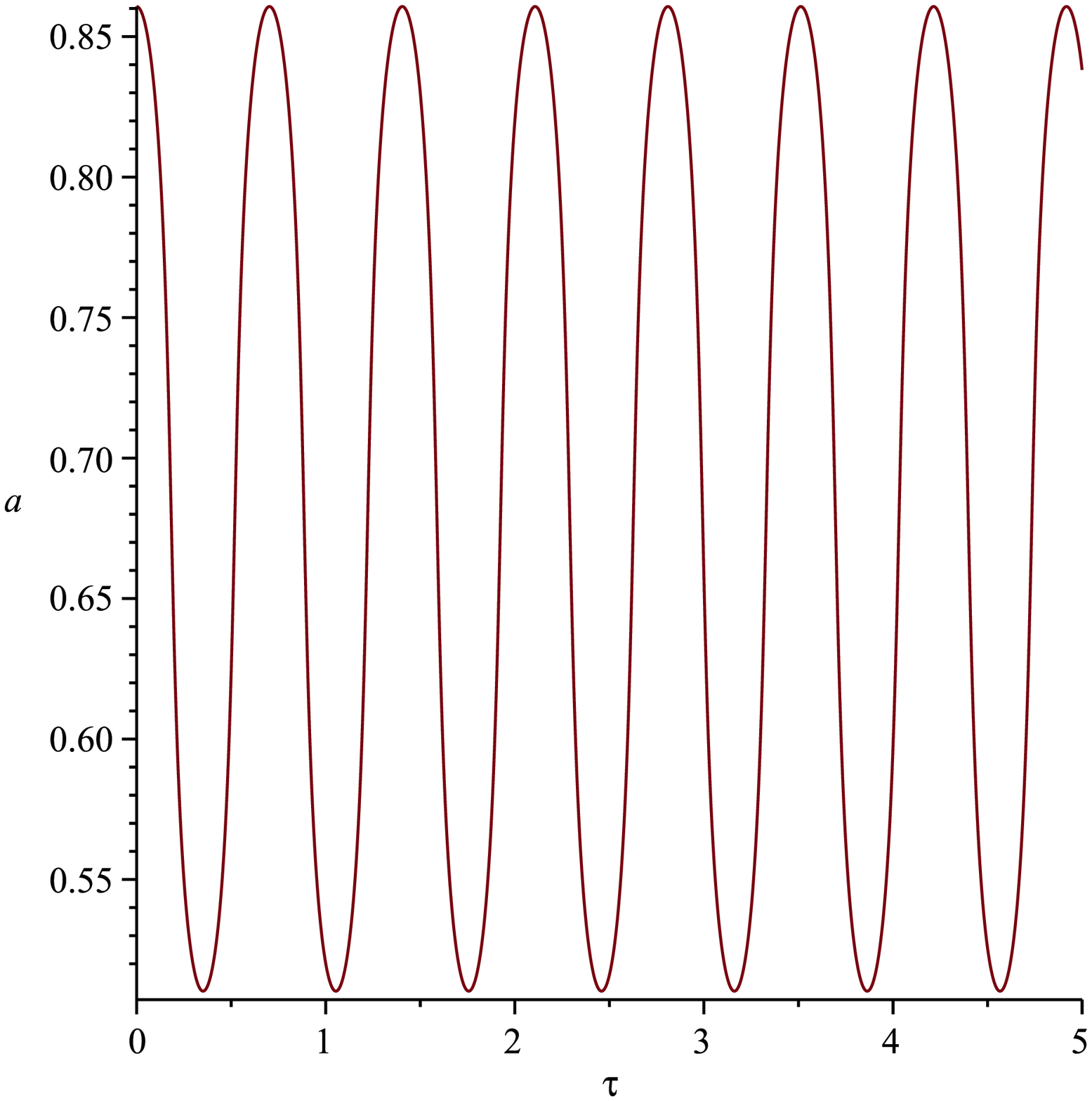}
\label{fig13a}
%}
%\quad
%\subfloat[Incerteza e valor esperado para $g_{c}=10$.]{
\includegraphics[scale=0.3]{%Gr=80-axt
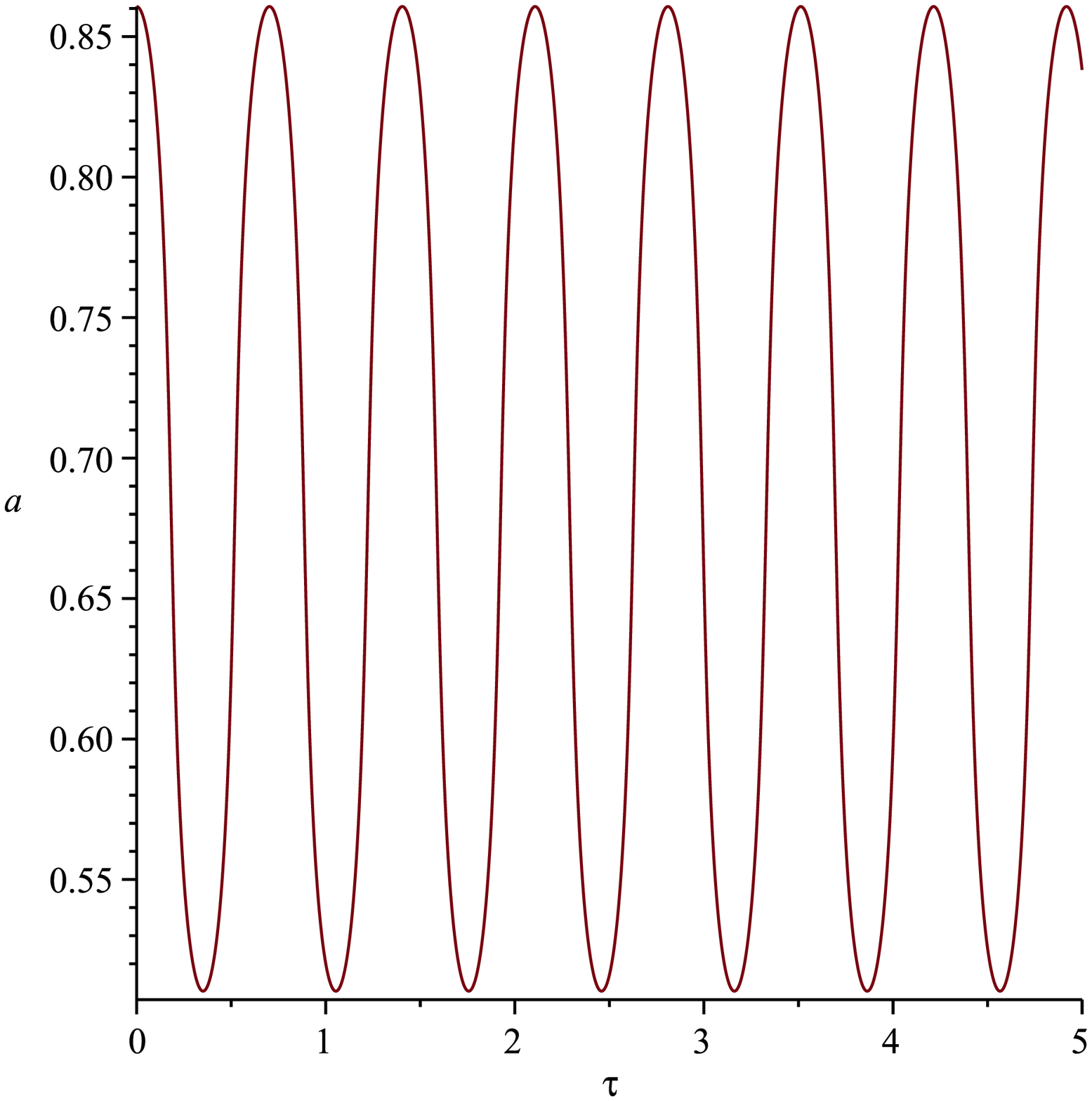}
\label{fig13b}
%}
\caption{Bohmian trajectories of $a$ for $N=2$, $g_s=-2$, $g_\Lambda=0$, $g_c=20$. Left: we have $g_r=40$. Right: we have $g_r=80$.}
\label{fig13}
\end{figure}

\subsubsection{Bohmian trajectories of $a$ as $g_s$ varies}

Solving equation (\ref{43}) for several different values of $g_s$, $N$ and various $\tau$ intervals, while keeping fixed the other HL parameters, 
we have observed the following behavior of the Bohmian  trajectories of $a$, as $g_s$ varies: 
($i$) the amplitude of oscillation and the number of oscillations of $a$, for a fixed $\tau$ interval, do not vary;
($ii$) the maximum value of $a$ increases as $g_s$ decreases. Such behavior agree with that of expected value $\left<a\right>$,
obtained in the previous section. This is exemplified by Figs. 8 and 14.

\begin{figure}[h!]
\centering
%\subfloat[Incerteza e valor esperado para $g_\Lambda=-1$.]{
\includegraphics[scale=0.3]{%Gs=-20-axt
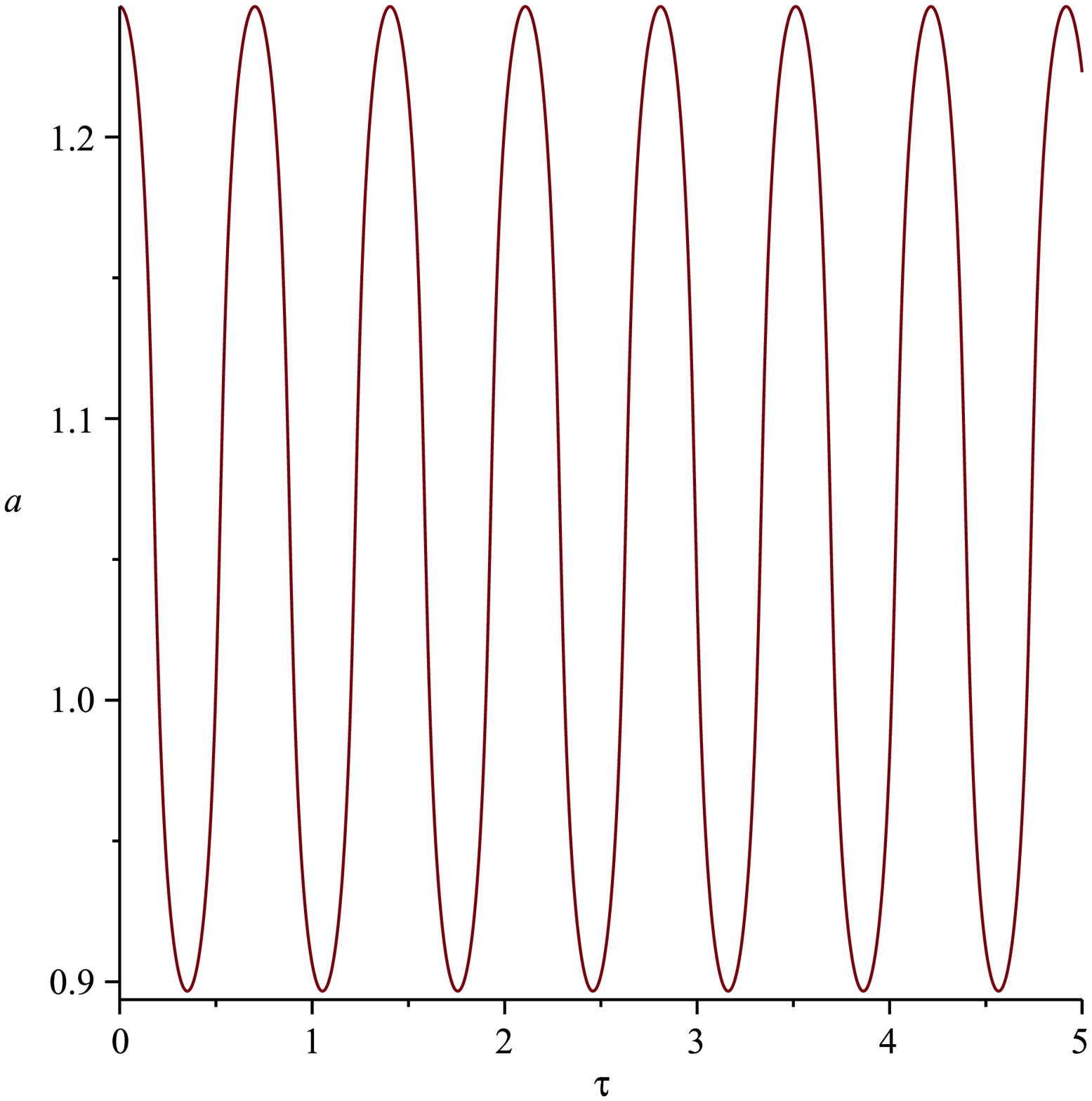}
\label{fig14a}
%}
%\quad
%\subfloat[Incerteza e valor esperado para $g_{c}=10$.]{
\includegraphics[scale=0.3]{%Gs=-120-axt
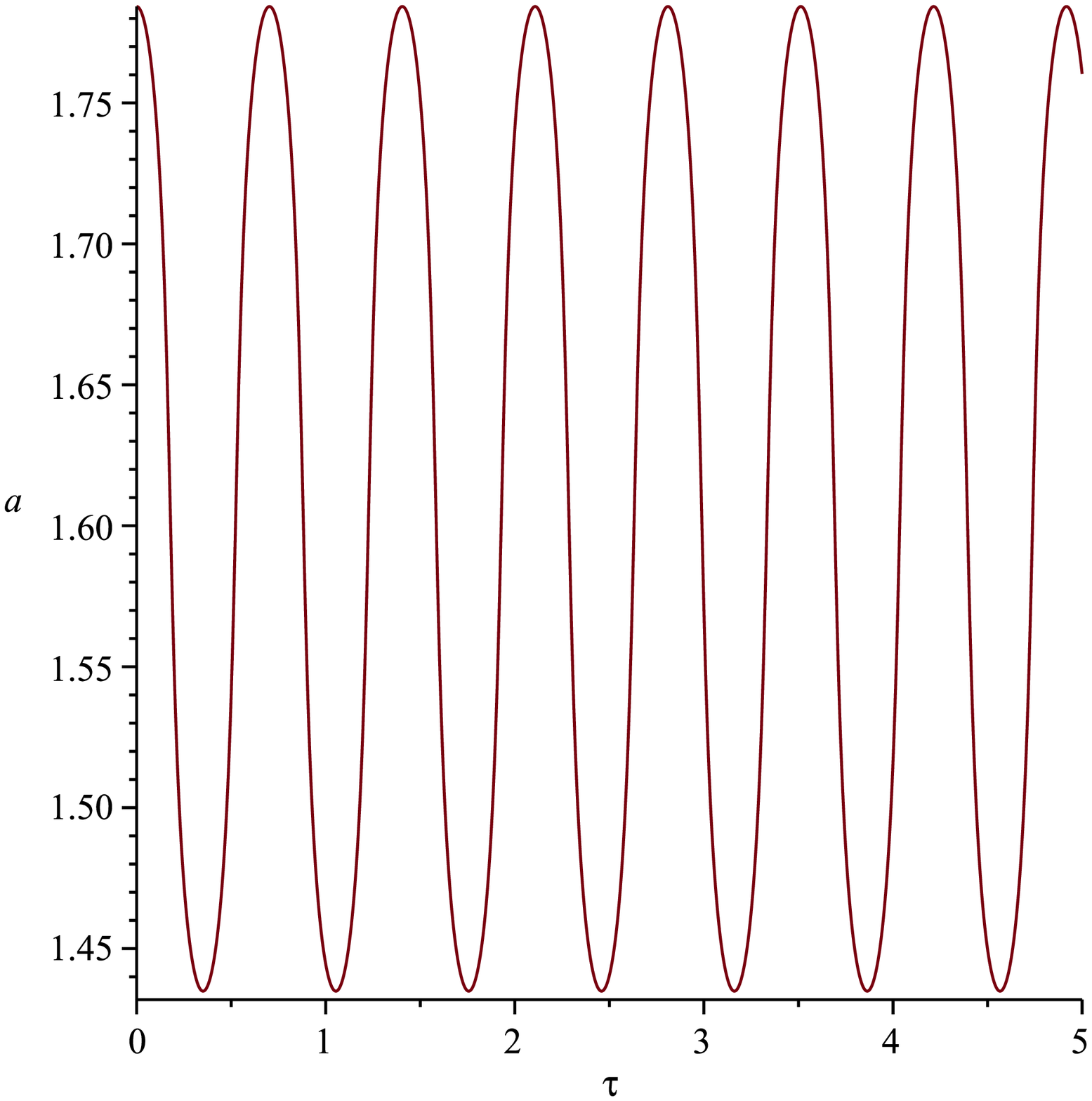}
\label{fig14b}
%}
\caption{Bohmian trajectories of $a$ for $N=2$, $g_r=120$, $g_\Lambda=0$, $g_c=20$. Left: we have $g_s=-20$. Right: we have $g_s=-120 $.}
\label{fig14}
\end{figure}

\subsection{The Bohmian quantum potential}

Observing the quantum potential $Q(a,\tau)$ in Eq. (\ref{42}) for the
present models, it is not difficult to understand why they are free from
singularities.
Here, together with the Bohmian trajectories of $a$, we
have also computed the potential $Q(a,\tau)$ for many different values
of $g_c$, $g_\Lambda$, $g_r$, $g_s$ and $N$. The calculations have been
made over each Bohmian trajectory of $a$. For all situations considered,
we obtained the same qualitative behavior for $Q(a,\tau)$.

If we compute the quantum potential $Q$ as a function of $\tau$ only, we see that it oscillates between maximum and minimum values. 
There are two different types of maximum values: the absolute maxima and the local maxima.
The absolute maxima values are greater than the local maxima values. 
The absolute maxima of $Q$ occur as the scale factor reaches its minimum values. Therefore, $Q(a,\tau)$ prevents $a(\tau)$ from ever vanishing.
On the other hand, the local maxima of $Q$ occur as the scale factor reaches its maximum values. 
In this way, the quantum potential prevents the scale factor from reaching infinite values. 
We have also computed $Q(a,\tau)$ as a function of $a$ only. 
In that case, for all considered values of $g_c$, $g_\Lambda$, $g_r$, $g_s$ and $N$, we have observe the same type of quantum potential
curve. The absolute and local maxima values of $Q$ can be clearly identified in that curve. 
Fig. 15 exemplifies the Bohmian quantum potential Eq. (\ref{42}) for the model with $g_{\Lambda}=-1$, $g_c=10$, $g_r=120$, $g_s=-2$ and $N=2$. 
In the left panel of Fig. 15, $Q$ is shown as a function of the time $\tau$; in the right panel, $Q$ is shown as a function of $a$. 
For a better understanding of the behavior of $Q$ it is important to observe the Bohmian trajectory of $a$
plotted on the left panel of Fig. 12. The same time interval used for $a$ in Fig. 12 has been used for $Q$ in Fig. 15, and the initial condition $a(\tau=0)$ for that trajectory was 
given by the expected value of $a$ at $\tau=0$, for the same model.

\begin{figure}[h!]
\centering
%\subfloat[Potencial quântico $\times$ tempo.]{
\includegraphics[scale=0.3]{%GL=-1-Qxt
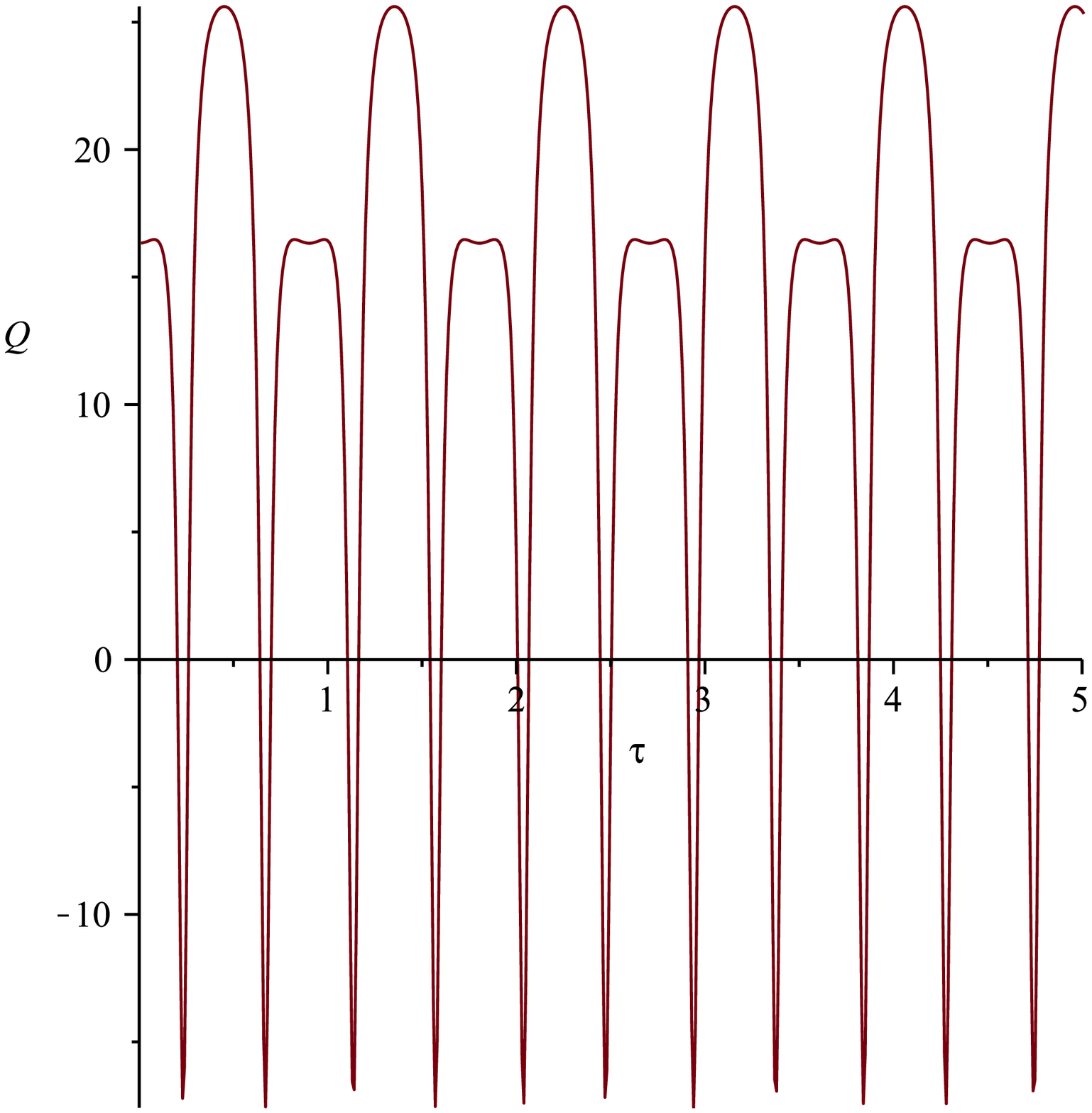}
\label{fig15a}
%}
%\quad
%\subfloat[Potencial quântico $\times$ fator de escala]{
\includegraphics[scale=0.3]{%GL=-1-Qxa
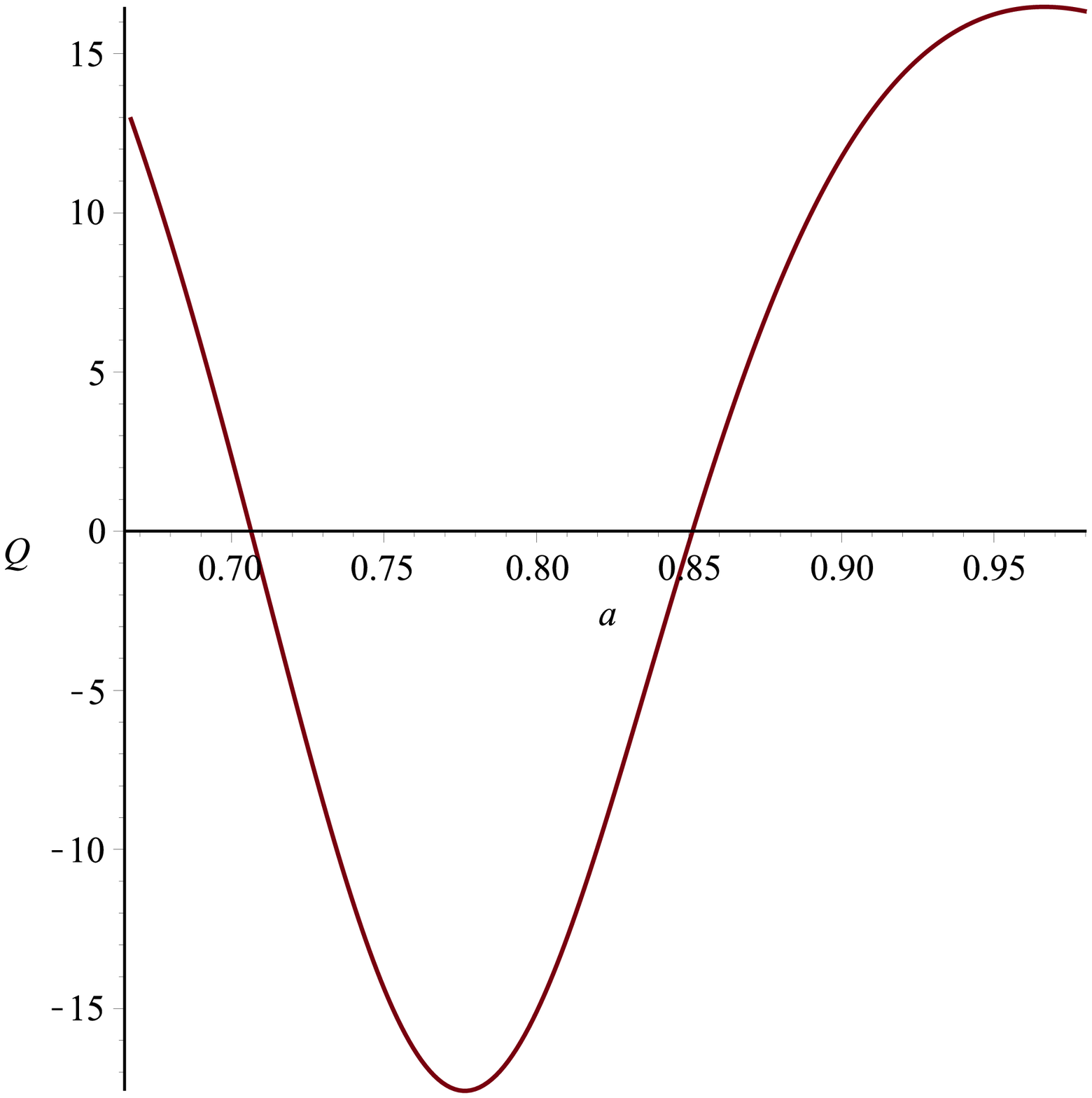}
\label{fig15b}
%}
\caption{Left: the quantum potential $Q$ as a function of the time $\tau$. Right: the quantum potential $Q$ as a function of the scale factor $a$. 
In both panels, we have $g_{\Lambda}=-1$, $N=2$, $g_c=10$, $g_r=120$ and $g_s=-2$.
The potential $Q$ shown has been computed along the Bohmian trajectory of $a$ presented on the left panel of Fig. 12.}
\label{fig15}
\end{figure}

\section{Conclusions}
\label{conclusions}

In the present paper, we canonically quantized a homogeneous and isotropic Ho\v{r}ava-Lifshitz 
cosmological model, with constant positive spatial sections and coupled to radiation. We considered 
the projectable version of that gravitational theory without the detailed balance condition. We used the
ADM formalism to write the gravitational Hamiltonian of the model and the Schutz variational
formalism to write the perfect fluid Hamiltonian. We obtained the Wheeler-DeWitt equation for the
model, which depends on several parameters coming from the HL theory. 
We studied the case of bounded solutions to the Wheeler-DeWitt equation, and the HL parameters have been chosen accordingly.

First, we have solved it using the {\it Many Worlds} interpretation of quantum mechanics. Using wavepackets computed with
the solutions to the Wheeler-DeWitt equation, we obtained the  expected value of the scalar factor $\left<a\right>$. 
We showed that this quantity oscillates between maximum and minimum values and never vanishes, indicating that the model is free from singularities, at the quantum level. 
We have also reinforced this indication by showing that if we subtract a standard deviation unit of $a$ from the expected value $\left<a\right>$, 
a positive value is still obtained. We have also studied how the  expected value of scale factor depends on each of the HL parameters and $N$. 

Then we have used the {\it DeBroglie-Bohm} interpretation of quantum mechanics. First, by computing the Bohmian trajectories of $a$,
for many different values of the HL parameters and $N$. We showed that $a$, for all those trajectories, oscillates between maximum and minimum
values and never vanish, in agreement with the behavior of the expected value of $a$. We were able to evaluate how those trajectories depend 
on the HL parameters and $N$ and compare the Bohmian trajectories of $a$ to the expected value $\left<a\right>$, showing that they agree for the corresponding models. 
Finally, we computed the quantum potential $Q$, for many different values of the HL parameters and $N$, showing how that quantity helps understanding 
why the scale factor never vanishes, in the present HL cosmological model.

\section*{Acknowledgments}

This study was financed in part by the Coordena\c{c}\~{a}o de Aperfei\c{c}oamento de Pessoal de N\'{i}vel Superior - Brasil (CAPES) - Finance Code 001.
L. G. Martins thanks CAPES for her scholarship.

\end{document}